\documentclass[preprint,superscriptaddress,showpacs,nofootinbib,aps]{revtex4}
 \usepackage[]{graphicx,subfigure}
 \usepackage{color}
 \usepackage{ulem}
 \def\be{\begin{equation}}
 \def\ee{\end{equation}}
 \def\ba{\begin{eqnarray}}
 \def\ea{\end{eqnarray}}
 \begin{document}
 \title{Spectator Scattering and Annihilation Contributions \\
        as a Solution to the ${\pi}K$ and ${\pi}{\pi}$ Puzzles \\
        within QCD Factorization Approach}
 \author{Qin Chang}
  \email{changqin@htu.edu.cn}
 \affiliation{Institute of Particle and Nuclear Physics,
              Henan Normal University, Xinxiang 453007, China}
 \affiliation{Institute of Particle Physics,
              Central China Normal University,
              Wuhan 430079, China}
 \author{Junfeng Sun}
 \email{sunjunfeng@htu.edu.cn}
 \affiliation{Institute of Particle and Nuclear Physics,
              Henan Normal University, Xinxiang 453007, China}
 \author{Yueling Yang}
 \email{yangyueling@htu.edu.cn}
 \affiliation{Institute of Particle and Nuclear Physics,
              Henan Normal University, Xinxiang 453007, China}
 \author{Xiaonan Li}
 \affiliation{Institute of Particle and Nuclear Physics,
              Henan Normal University, Xinxiang 453007, China}
 \begin{abstract} 
 The large branching ratios for pure annihilation $\bar{B}_s^0$ $\to$ 
 $\pi^+ \pi^-$ and $\bar{B}_d^0$ $\to$ $K^+ K^-$ decays
 reported by CDF and LHCb collaborations recently
 and the so-called ${\pi}K$ and ${\pi}{\pi}$ puzzles
 indicate that spectator scattering and annihilation
 contributions are important to the penguin-dominated,
 color-suppressed tree dominated, and pure annihilation
 nonleptonic $B$ decays.
 Combining the available experimental data for $B_{u,d}$
 ${\to}$ $\pi \pi$, ${\pi}K$ and $K \bar{K}$ decays, we do 
 a global fit on the spectator scattering and annihilation parameters
$X_H({\rho}_H$, ${\phi}_H)$, $X_A^i({\rho}_A^{i},{\phi}_A^{i})$ and
 $X_A^f({\rho}_A^{f},{\phi}_A^{f})$, which are used to
 parameterize the endpoint singularity in amplitudes
 of spectator scattering, nonfactorizable and
 factorizable annihilation topologies within the QCD
 factorization framework, in three scenarios for different purpose.
 Numerically, in scenario II, we get $({\rho}_A^{i},{\phi}_A^{i}[^{\circ}])=(2.88^{+1.52}_{-1.30},-103^{+33}_{-40})$
 and $({\rho}_A^{f},{\phi}_A^{f}[^{\circ}])=(1.21^{+0.22}_{-0.25},-40^{+12}_{-8})$ at the $68\%$
 confidence level, which are mainly demanded by resolving 
 ${\pi}K$ puzzle and confirm the presupposition that $X_A^i\neq X_A^f$.
 In addition, correspondingly, the $B$-meson wave function
 parameter $\lambda_B$ is also fitted to be $0.18^{+0.11}_{-0.08}\, MeV$, 
 which plays an important role for resolving both ${\pi}K$ and $\pi\pi$ puzzles. 
 With the fitted parameters, the QCDF
 results for observables of $B_{u,d}$ $\to$ $\pi \pi$,
 $\pi K$ and $K \bar{K}$ decays are in good agreement
 with experimental measurements.
 Much more experimental and theoretical efforts are
 expected to understand the underlying QCD dynamics
 of spectator scattering and annihilation contributions.
 \end{abstract}
 \maketitle

 \section{Introduction}
 \label{sec01}
 Charmless hadronic $B$-meson decays provide a fertile
 ground for testing the Standard Model (SM) and
 exploring the source of $CP$ violation, which attract
 much attention in the past years.
 Thanks to the fruitful accomplishment of BABAR and Belle,
 the constraints on the sides and interior angles
 of the unitarity triangle significantly reduce
 the allowed ranges of some of the CKM elements,
 and many rare $B$ decays are well measured.
 With the successful running of LHC and
 the advent of Belle II at SuperKEKB,
 heavy flavour physics has entered a new
 exciting era and more $B$ decay modes
 will be measured precisely soon.

 Recently, the evidence of pure annihilation decays
 $\bar{B}_{s}^{0}$ ${\to}$ ${\pi}^{+}{\pi}^{-}$ and
 $\bar{B}_{d}^{0}$ ${\to}$ $K^{+}K^{-}$ are firstly
 reported by CDF Collaboration \cite{CDFanni}, and
 soon confirmed by LHCb Collaboration \cite{LHCbanni}.
 The Heavy Flavor Averaging Group (HFAG) presents
 their branching ratios \cite{HFAG}
  \begin{equation}
 {\cal B}(\bar{B}_{s}^{0}{\to}{\pi}^{+}{\pi}^{-})
  =(0.73{\pm}0.14){\times}10^{-6}
  \label{HFAGpipi},
  \end{equation}
  \begin{equation}
 {\cal B}(\bar{B}_{d}^{0}{\to}K^{+}K^{-})
  =(0.12{\pm}0.05){\times}10^{-6}
  \label{HFAGKK}.
  \end{equation}
 Such results, if confirmed, imply unexpectedly large
 annihilation contributions in $B$ decays and
 significant flavour symmetry breaking effects
 between the annihilation amplitudes of $B_{u,d}$
 and $B_{s}$ decays, which attract much attention
 recently, for instance Refs.
 \cite{zhu1,zhu2,chang1,xiao1}.

 Theoretically,  as noticed already in Refs. \cite{pqcd,relaRef,du1,Beneke2}, 
 even though the annihilation contributions are formally 
 $\Lambda_{QCD}/m_b$ power suppressed, they are very important 
 and indispensable for charmless $B$ decays.
 By introducing the parton transverse momentum and
 the Sudakov factor to regulate the endpoint divergence,
 there is a large complex annihilation contribution within
 the perturbative QCD (pQCD) approach \cite{pqcd,relaRef}.
 The latest renewed pQCD estimations\footnotemark[1]
 \footnotetext[1]{The first three uncertainties come
 from meson wave functions, the last one is from
 the CKM factors.}
 ${\cal B}(\bar{B}_s^0 \to \pi^+ \pi^-)$ $=$
 $(5.10^{+1.96+0.25+1.05+0.29}_{-1.68-0.19-0.83-0.20}) \times 10^{-7}$
 and ${\cal B}(\bar{B}_d^0 \to K^+ K^-)$ $=$
 $(1.56^{+0.44+0.23+0.22+0.13}_{-0.42-0.22-0.19-0.09}) \times 10^{-7}$
 \cite{xiao1} give an appropriate account of the CDF
 and LHCb measurements within uncertainties.
 In the QCD factorization (QCDF) framework
 \cite{Beneke1}, the endpoint divergence in
 annihilation amplitudes is usually
 parameterized by $X_{A}(\rho_A,\phi_A)$ (see Eq.(\ref{XA})).
 The parameters $\rho_A$ $\sim$ $1$
 and $\phi_A$ $\sim$ $-55^{\circ}$
 (scenario S4) \cite{Beneke2} are
 adopted conservatively in evaluating the
 amplitudes of $B$ $\to$ $PP$ decays,
 which lead to the predictions\footnotemark[2]
 \footnotetext[2]{The second uncertainty comes
 from parameters ${\rho}_{A,H}$ and ${\phi}_{A,H}$
 of annihilation and spectator contributions.}
 ${\cal B}(\bar{B}_s^0 \to \pi^+ \pi^-)$
 $=$ $(0.26^{+0.00+0.10}_{-0.00-0.09}) \times 10^{-6}$
 and ${\cal B}(\bar{B}_d^0 \to K^+ K^-)$
 $=$ $(0.10^{+0.03+0.03}_{-0.02-0.03}) \times 10^{-6}$
 \cite{Cheng2}.
 It is obvious that the QCDF prediction of
 ${\cal B}(\bar{B}_d^0 \to K^+ K^-)$ agrees well with
 the data Eq.(\ref{HFAGKK}),
 but the one of ${\cal B}(\bar{B}_s^0 \to \pi^+ \pi^-)$ is
 much smaller than the present experimental
 measurement Eq.(\ref{HFAGpipi}).
 This discrepancy kindles the passions of
 restudy on annihilation contributions
 \cite{zhu1,zhu2,chang1}.

 At present, there are two major issues among the
 well-concerning focus on the annihilation contributions
 within the QCDF framework, one is whether $X_A(\rho_A,\phi_A)$ is universal for $B$ decays,
 and the other is what its value should be.
 As to the first issue, there is no an imperative reason for the
 annihilation parameters $\rho_A$ and $\phi_A$ to
 be the same for different $B_{u,d,s}$ decays,
 even for different annihilation topologies,
 although they were usually taken to be universal in
 the previous numerical calculation for simplicity \cite{du1,Beneke2}.
 Phenomenologically, it is almost
 impossible to account for all of the well-measured
 two-body charmless $B$ decays with the universal
 values of $\rho_A$ and $\phi_A$ based on
 the QCDF approach \cite{zhu2,chang1,Beneke2,Cheng2}.
 In addition, the pQCD study on $B$ meson decays
 also indicate that the annihilation parameters
 $\rho_A$ and $\phi_A$ should be process-dependent.
 In fact, in the practical QCDF application to the
 $B$ ${\to}$ $PP$, $PV$ decays (where $P$ and $V$
 denote the light pseudoscalar and vector $SU(3)$
 meson nonet, respectively),
 the non-universal values of annihilation phase
 $\phi_A$ with respect to PP and PV final states are favored
 (scenario S4) \cite{Beneke2};
 the process-dependent values of $\rho_A$ and
 $\phi_A$ are given based on an educated guess
 \cite{Cheng2,Cheng1} or the comparison with the updated
 measurements \cite{chang1};
 the flavour-dependent values of $\rho_A$ and
 $\phi_A$ are suggested recently in the
 nonfactorizable annihilation contributions
 \cite{zhu2}.
 In principle the value of $\rho_A$ and $\phi_A$
 should differ from each other for different topologies
 with different flavours, but we hope that the QCDF
 approach can accommodate and predict much more hadronic
 $B$ decays with less input parameters.
 So much attention in phenomenological analysis on the
 weak annihilation $B$ decays is devoted to what the
 appropriate values of the parameters $\rho_A$ and
 $\phi_A$ should be. This is the second issue.
 In principle, a large value of $\rho_A$ is unexpected
 by the power counting rules and the self-consistency
 validation within the QCDF framework.
 The original proposal is that $\rho_A$ ${\leq}$ $1$
 and an arbitrary strong interaction phase $\phi_A$
 are universal for all decay processes, and that
 a fine-tuning of the phase $\phi_A$ is required
 to be reconciled with experimental data when $\rho_A$
 is significantly larger than 1 \cite{Beneke2}.
 The recent study on the annihilation contributions
 show that $\rho_A$ $>$ $2$ and ${\vert}\phi_A{\vert}$
 $\geq$ $30^{\circ}$ are acceptable, even necessary,
 to reproduce the data for some two-body nonleptonic
 $B_{u,d,s}$ decay modes \cite{zhu2,chang1}.
 In this paper, we will perform a fitting on the
 parameters $\rho_A$ and $\phi_A$ by considering
 $B$ ${\to}$ ${\pi}{\pi}$, ${\pi}K$ and $K \bar{K}$
 decay modes, on one hand, to investigate the strength
 of annihilation contribution,
 on the other hand, to study their effects on the
 anomalies in $B$ physics, such as the
 well-known ${\pi}K$ and ${\pi}{\pi}$ puzzles.

 The so-called ${\pi}K$ puzzle is reflected by
 the difference between the direct $CP$ asymmetries
 for $B^{-}$ ${\to}$ $K^{-}\pi^{0}$ and
 $\bar{B}^{0}$ ${\to}$ $K^{-}\pi^{+}$ decays.
 With the up-to-date HFAG results \cite{HFAG},
 we get
 \begin{equation}
 \Delta A \equiv A_{CP}(B^{-} {\to} K^{-} {\pi}^{0})
   - A_{CP}(\bar{B}^{0} \to K^{-} \pi^{+})
   = (12.2 \pm 2.2) \%
 \label{acppi},
 \end{equation}
 which differs from zero by about $5.5\sigma$.
 However, the direct $CP$ asymmetries of
 $A_{CP}(B^{-} \to K^{-} \pi^{0})$ and
 $A_{CP}(\bar{B}^{0} \to K^{-} \pi^{+})$
 are expected to be approximately equal
 with the isospin symmetry in the SM,
 numerically for instance $\Delta A \sim 0.5 \%$
 in the S4 scenario of QCDF \cite{Beneke2}.

 The so-called ${\pi}{\pi}$ puzzle is reflected by
 the following two ratios of the $CP$-averaged
 branching fractions \cite{pipipuz}:
 \begin{equation}
 R_{+-}^{\pi \pi}
 \equiv 2 \Big[
 \frac{ {\cal B}(B^{-} \to \pi^{-} \pi^0) }
      { {\cal B}(\bar{B}^{0} \to \pi^{+} \pi^{-}) }
 \Big]
 \frac{ \tau_{B^0} }{ \tau_{B^+} },
 \qquad
 R_{00}^{\pi \pi}
 \equiv  2 \Big[
 \frac{ {\cal B}( \bar{B}^{0} \to \pi^0 \pi^0) }
      { {\cal B}( \bar{B}^{0} \to \pi^{+} \pi^{-}) }\Big]
 \label{pipipuzzle}.
 \end{equation}
 It is generally expected that branching ratio
 ${\cal B}(\bar{B}^{0} \to \pi^+ \pi^-) \gtrsim
  {\cal B}(B^{-} \to \pi^{-} \pi^0)$ and
 ${\cal B}(\bar{B}^{0} \to \pi^+ \pi^-) \gg
  {\cal B}(\bar{B}^{0} \to \pi^0 \pi^0)$
 within the SM.
 To date, the agreement of $R_{+-}^{\pi \pi}$
 between the S4 scenario QCDF
 $R_{+-}^{\pi \pi}(\text{QCDF})$ $=$ $1.83$
 \cite{Beneke2} and the refined experimental data
 $R_{+-}^{\pi \pi}(\text{Exp.})$ $=$ $1.99 \pm 0.15$
 \cite{HFAG} can be achieved consistently within
 experimental error,
 while the discrepancy in $R_{00}^{\pi \pi}$
 between the S4 scenario QCDF
 $R_{00}^{\pi \pi}(\text{QCDF})$ $=$ $0.27$
 (where theoretical uncertainties are unenclosed)
 \cite{Beneke2} and the progressive experimental
 data $R_{+-}^{\pi \pi}(\text{Exp.})$ $=$
 $1.99 \pm 0.15$ \cite{HFAG} is unexpectedly
 large.

 It is claimed \cite{pipipuz} that the so-called
 ${\pi}{\pi}$ puzzle could be accommodated by the
 nonfactorizable contributions in SM.
 It is argued \cite{Cheng1,pipipuz} that to solve
 the so-called ${\pi}K$ puzzle, a large complex
 color-suppressed tree amplitude $C^{\prime}$
 or a large complex electroweak penguin contribution
 $P_{\rm EW}^{\prime}$ or a combination of them
 are essential.
 An enhanced complex $P_{\rm EW}^{\prime}$ with
 a nontrivial strong phase can be obtained from
 new physics effects \cite{pipipuz}.
 To get a large complex $C^{\prime}$, one can
 resort to spectator scattering and final state
 interactions \cite{Cheng1,Cheng2}.
 Recently, the annihilation amplitudes with
 large parameters $\rho_A$ is suggested to
 conciliate the recent measurements Eq.(\ref{HFAGpipi})
 and Eq.(\ref{HFAGKK}),
 so surprisingly, the ${\pi}K$ puzzle is also
 resolved simultaneously \cite{zhu2}.
 Theoretically, the power corrections, such as
 spectator scattering at the twist-3 order and
 annihilation amplitudes, are important to
 account for the large branching ratios and
 $CP$ asymmetries of penguin-dominated and/or
 color-suppressed tree-dominated $B$ decays.
 So, before claiming a new physics signal, it is
 essential to examine whether power corrections
 could retrieve ``problematic'' deviations from
 the SM expectations.
 Interestingly, our study show that with
 appropriate parameters, the annihilation and spectator
 scattering contributions could provide some possible
 solutions to the $\pi K$ and $\pi \pi$ puzzles.

 Our paper is organized as following.
 In section \ref{sec02}, we give a brief overview
 of the hard spectator and annihilation calculations
 and recent studies within QCDF.
 In section \ref{sec03}, focusing on $\pi K$ and
 $\pi \pi$ puzzles, the effects of spectator scattering
 and annihilation contributions on $B$ $\to$ $\pi \pi$,
 $\pi K$ and $K \bar{K}$ decays are studied
 in detail in blue{three} scenarios.
 In each  scenario, a fitting on relevant parameters
 are performed.
 Our conclusions are summarized in section \ref{sec04}.
 Appendix \ref{app01} recapitulates the building blocks
 of annihilation and spectator scattering amplitudes.
 The input parameters and our fitting approach
 are given in Appendix \ref{app02} and \ref{app03},
 respectively.

 \section{Brief Review of Spectator Scattering and
   Annihilation Amplitudes within QCDF}
 \label{sec02}
 The effective Hamiltonian for nonleptonic $B$ weak
 decays is \cite{Buchalla:1996vs}
 \begin{eqnarray}
 {\cal H}_{\rm eff} &=&
  \frac{G_F}{\sqrt{2}}
  \sum\limits_{p,q}
  V_{pb} V_{pq}^{\ast}
  \Big\{
  \sum\limits_{i=1}^{10}
   C_i O_i + C_{7 \gamma} O_{7 \gamma}
   + C_{8g} O_{8g} \Big\}
   + {\rm h.c.}
 \label{eq:eff},
 \end{eqnarray}
 where $V_{pb} V_{pq}^{\ast}$ ($p$ $=$ $u$, $c$
 and $q$ $=$ $d$, $s$)
 is the product of the Cabibbo-Kobayashi-Maskawa
 (CKM) matrix elements;
 $C_{i}$ is the Wilson coefficient corresponding to
 the local four-quark operator $O_i$;
 $O_{7 \gamma}$ and $O_{8g}$ are the electromagnetic
 and chromomagnetic dipole operators.

 \begin{figure}[t]
 \begin{center}
 \subfigure[]{\includegraphics[width=0.22\textwidth]{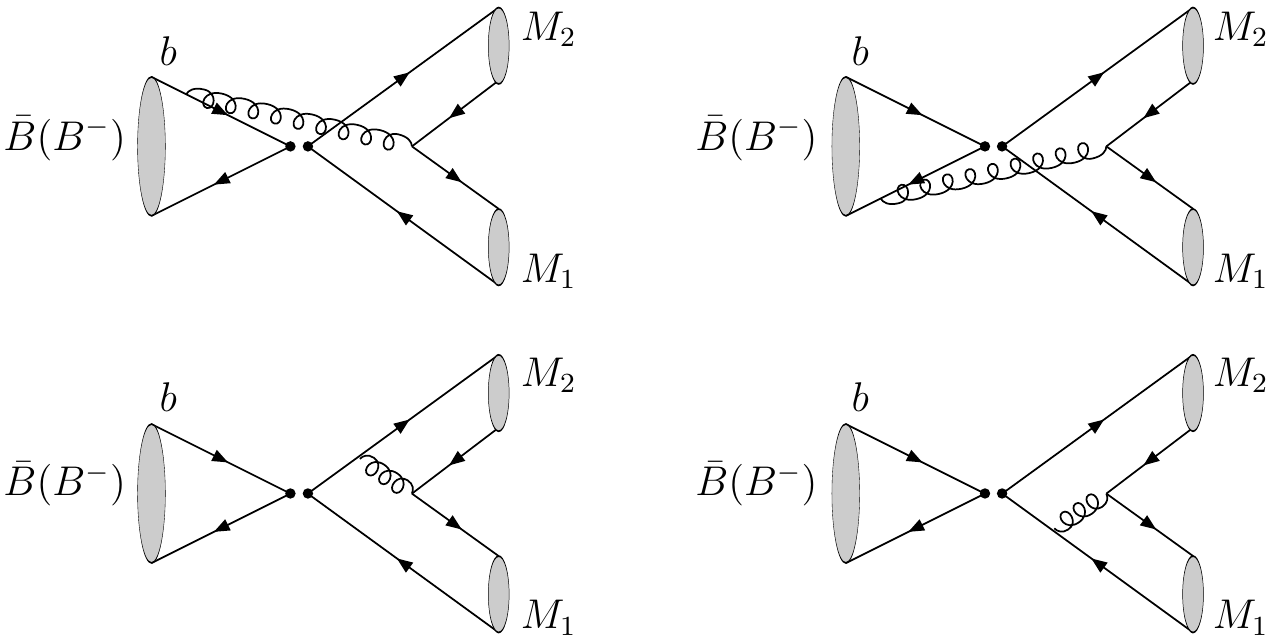}}
 \quad
 \subfigure[]{\includegraphics[width=0.22\textwidth]{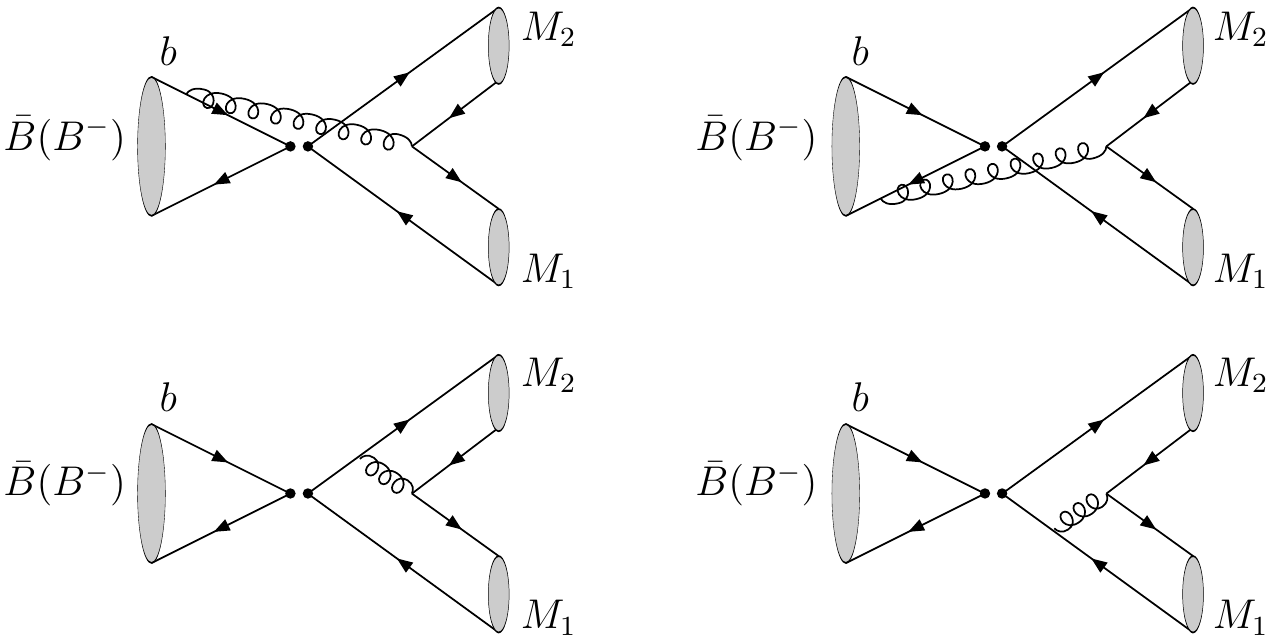}}
 \quad
 \subfigure[]{\includegraphics[width=0.22\textwidth]{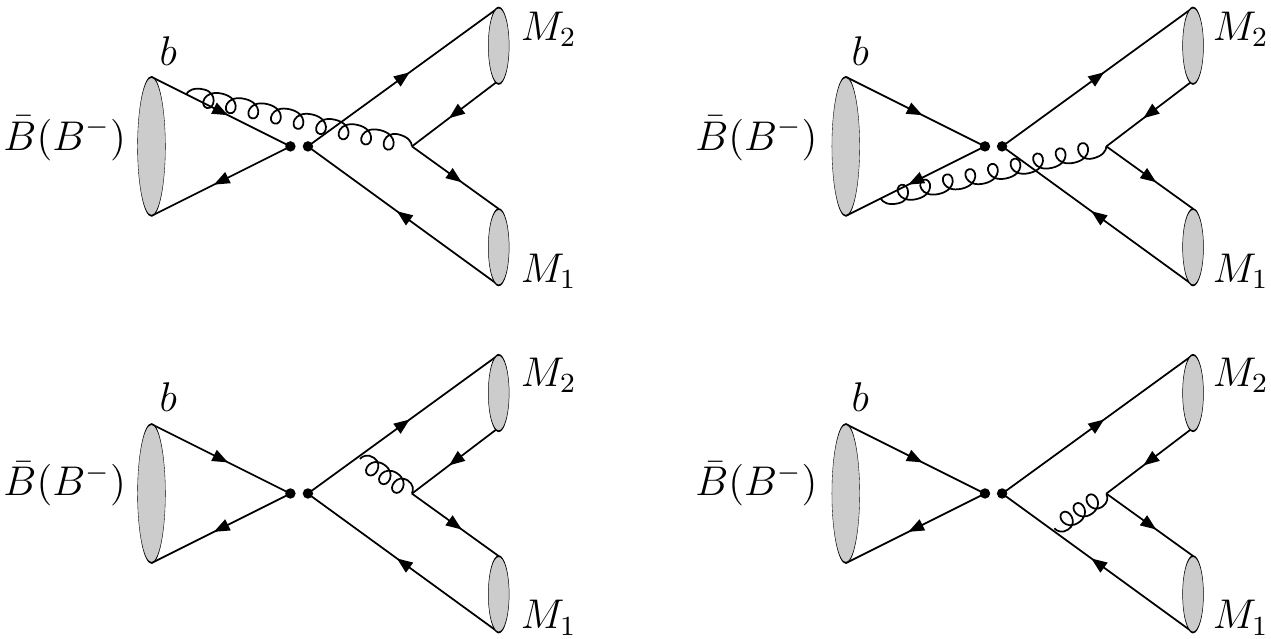}}
 \quad
 \subfigure[]{\includegraphics[width=0.22\textwidth]{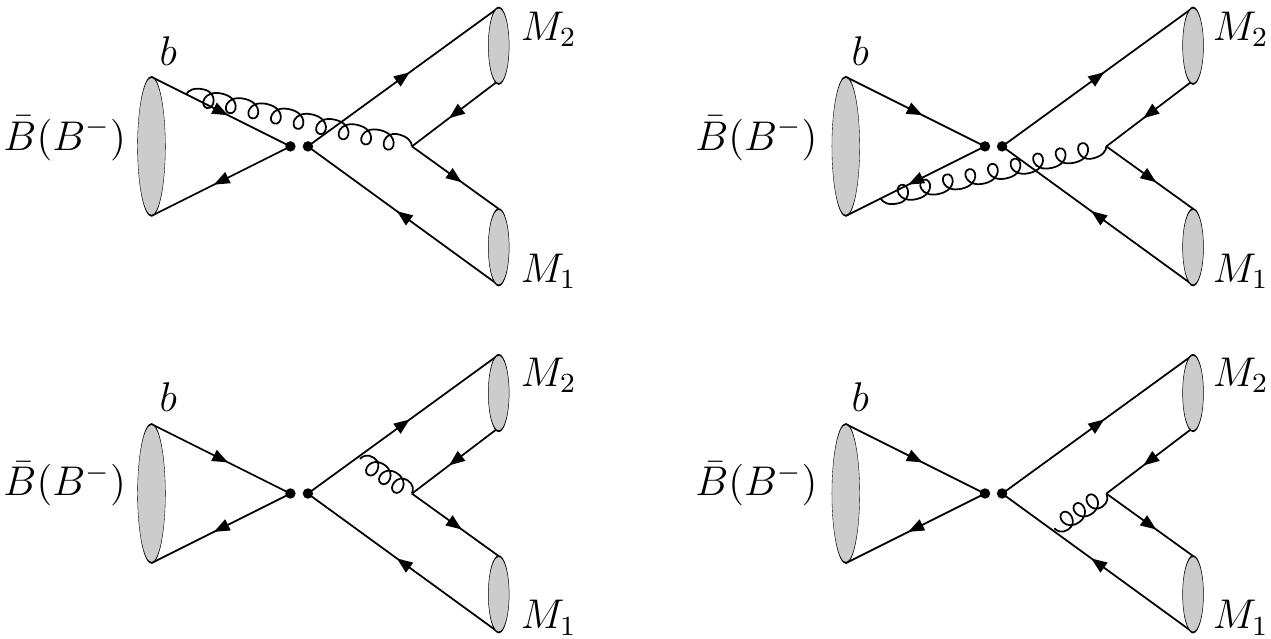}} \\
 \subfigure[]{\includegraphics[width=0.22\textwidth]{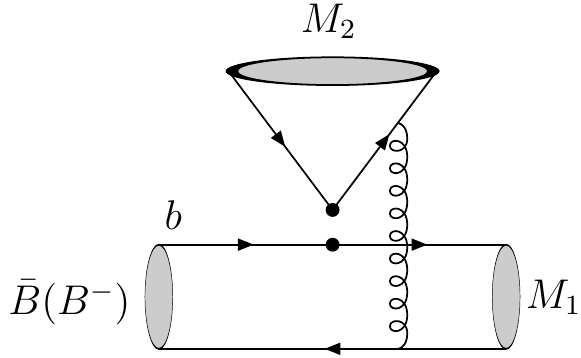}}
 \quad
 \subfigure[]{\includegraphics[width=0.22\textwidth]{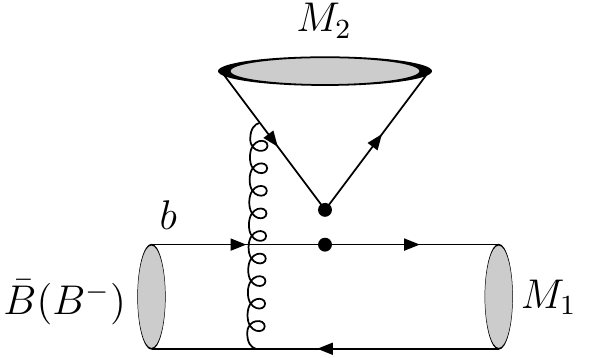}}
 \caption{
 The lowest order diagrams of weak annihilation (a-d) and
 spectator scattering (e,f).}
 \label{diag}
 \end{center}
 \end{figure}

 With the effective Hamiltonian Eq.(\ref{eq:eff}),
 the QCDF method has been fully developed and extensively
 employed to calculate the hadronic two-body $B$ decays,
 for example, see \cite{du1,Beneke1,Beneke2,Cheng2}.
 The spectator scattering and annihilation amplitudes
 (see Fig.\ref{diag}) are expressed as the convolution of
 scattering functions with the light-cone wave functions
 of the participating mesons \cite{Beneke1,Beneke2}.
 The explicit expressions for the basic building blocks
 of spectator scattering and annihilation amplitudes have
 been given by Ref. \cite{Beneke2}, which are also listed
 in the appendix \ref{app01} for convenience.
 With the asymptotic light-cone distribution amplitudes,
 the building blocks for annihilation amplitudes of
 Eq.(\ref{ai1}-\ref{af3}) could be simplified
 as \cite{Beneke2}
 \begin{eqnarray}
  A_1^i & \simeq & A_2^i \simeq
  2 \pi \alpha_s \Big[ 9\,\Big( X_A - 4 + \frac{\pi^2}{3} \Big)
 + r_\chi^{M_1} r_\chi^{M_2} X_A^2 \Big]
 \label{xai12}, \\
  A_3^i & \simeq &
  6 \pi \alpha_s \big(r_\chi^{M_1} - r_\chi^{M_2} \big)
  \Big( X_A^2 - 2 X_A + \frac{\pi^2}{3} \Big)
 \label{xai3}, \\
  A_3^f & \simeq &
  6 \pi \alpha_s ( r_\chi^{M_1} + r_\chi^{M_2} )
  (2 X_A^2 - X_A)
 \label{xaf3},
 \end{eqnarray}
 where the superscripts $i$ (or $f$) refers to gluon emission
 from the initial (or final) state quarks, respectively (see
 Fig.\ref{diag}).
 For the $\pi \pi$, $\pi K$ and $K \bar{K}$ final-state,
 $A_3^i$ is numerically negligible due to $r_\chi^{M_1}$
 $\simeq$ $r_\chi^{M_2}$.
 The model-dependent parameter $X_A$ is used to estimate
 the endpoint contributions, and expressed as
 \begin{equation}
 \int_0^1 \frac{dx}{x} \to
 X_A = (1+ \rho_A e^{i\phi_A})
 \ln \frac{m_B}{\Lambda_h}
 \label{XA},
 \end{equation}
 where $\Lambda_h$ $=$ $0.5$ GeV.
 For spectator scattering contributions, the calculation
 of twist-3 distribution amplitudes also suffers from
 endpoint divergence, which is usually dealt with the
 same manner as Eq.(\ref{XA}) and labelled by $X_H$ \cite{Beneke2}.
 Moreover, a quantity $\lambda_B$ is used to
 parameterize our ignorance about $B$-meson
 distribution amplitude [see Eq.(\ref{hardblock})]
 through \cite{Beneke2}
 \begin{equation}
 \int_0^1 \frac{ d \xi }{\xi} \Phi_B(\xi)
 \ \equiv \ \frac{m_B}{\lambda_B}
 \label{lamdef}.
 \end{equation}

 The QCDF approach itself cannot give information or/and
 constraint on the phenomenological parameters of $X_A$,
 $X_H$ and ${\lambda_B}$.
 These parameters should be determined from experimental
 data.
 To conform with measurements of nonleptonic $B$ ${\to}$
 $PP$ decays, we will adopt a similar method used in
 Ref.\cite{zhu2} to deal with the contributions from
 weak annihilation and spectator scattering.
 Focusing on the flavor dependence, without consideration
 of theoretical uncertainties, annihilation contributions
 are reevaluated in detail \cite{zhu2}
 to explain the ${\pi}K$ puzzle and the recent
 measurements on pure annihilation decays
 $\bar{B}_{s}^{0}$ ${\to}$ ${\pi}^{+}{\pi}^{-}$ and
 $\bar{B}_{d}^{0}$ ${\to}$ $K^{+}K^{-}$
 [see Eq.(\ref{HFAGpipi},\ref{HFAGKK})].
 The authors of Ref. \cite{zhu2} find that the flavour
 symmetry breaking effects should be carefully considered
 for $B_{u,d,s}$ decays, and suggest that
 the parameters of $\rho_A$ and $\phi_A$ in
 nonfactorizable annihilation topologies $A^{i}_{k}$
 [see Eq.(\ref{xai12},\ref{xai3})] should be different
 from those in factorizable annihilation topologies
 $A^{f}_{k}$ [see Eq.(\ref{xaf3})].
 (1)
 For factorizable annihilation topologies, i.e.,
 the gluon emission from the final states Fig.\ref{diag}(c,d),
 the flavor symmetry breaking effects are embodied in
 the decay constants, because the asymptotic light-cone
 distribution amplitudes of final states are the same.
 In addition, all decay constants have been factorized
 outside from the hadronic matrix elements of factorizable
 annihilation topologies.
 So $A^{f}_k$ is independent of the initial state,
 and is the same for $B_{u,d,s}$ annihilation decays to
 two light pseudoscalar mesons, that is to say,
 $\rho^f_A$ and $\phi^f_A$ should be universal
 for $B_{u,d,s}$ $\to$ $PP$ decays.
 (2)
 For nonfactorizable annihilation topologies,
 i.e., the gluon emission from the initial $B$ meson
 Fig.\ref{diag}(a,b), besides the factorized decay
 constants and the same asymptotic light-cone
 distribution amplitudes, $B$ meson wave functions
 $\Phi_{B}(\xi)$ are involved in the convolution
 integrals of hadronic matrix elements.
 Hence, $A^{i}_k$ should depend on the initial state
 and be different for $B_{u,d}$ from $B_{s}$ meosn
 due to flavor symmetry breaking effects, i.e.,
 parameters of $\rho^i_A$ and $\phi^i_A$ should be
 non-universal for $B_s$ and $B_{u,d}$ meson decays,
 and be different from parameters of
 $\rho^f_A$ and $\phi^f_A$ for $A^{f}_k$.
 In fact, the symmetry breaking effects have been
 considered in pervious QCDF study on two-body hadronic
 $B$ decays \cite{Cheng1,Cheng2,Cheng3,Beneke2,chang1},
 but with parameters of $\rho^f_A$ $=$ $\rho^i_A$
 and $\phi^f_A$ $=$ $\phi^i_A$.
 So, it is essential to systematically reevaluate
 factorizable and nonfactorizable annihilation
 contributions and preform a global fit on the
 annihilation parameters with the current
 available experimental data.
 In this paper, we will pay much attention to
 $B_{u,d}$ ${\to}$ $KK$, ${\pi}K$, ${\pi}{\pi}$
 decays and the aforementioned ${\pi}K$, ${\pi}{\pi}$
 puzzles with a distinction between
 ($\rho^f_A$, $\phi^f_A$) and ($\rho^i_A$, $\phi^i_A$),
 i.e., $X_A^i$ $\neq$ $X_A^f$.

 As aforesaid \cite{Cheng1,pipipuz},
 the nonfactorizable spectator scattering amplitudes
 contribute to a large complex $C^{\prime}$, which is
 important to resolve the ${\pi}K$, ${\pi}{\pi}$ puzzles.
 From the building block Eq.(\ref{hardblock}),
 it can be easily seen that $B$ meson wave functions
 $\Phi_{B}(\xi)$ appear in the spectator scattering
 amplitudes.
 Therefore, the symmetry breaking effects should
 also be considered for the quantity $X_H$ that
 is introduced to parameterize the endpoint
 singularity in the twist-3 level spectator
 scattering corrections.
 Similar to $X_A^i$, the quantity $X_H$ is related to the 
 topologies that gluon emit from the initial $B$ meson. 
 So, for simplicity, the approximation $X_H$ $=$ $X_A^i$ 
 is assumed in our coming numerical evaluation (scenarios I and II, 
 see the next section for detail).
 Of course, this approximation is neither 
 based on solid ground or from some underlying principle,
 and should be carefully studied and deserve much research.
 In fact, our coming phenomenological study (scenarios III) 
 shows that the approximation $X_H$ $=$ $X_A^i$
 is allowable with the up-to-date measurement on $B_{u,d}$ 
 ${\to}$ $KK$, ${\pi}K$, ${\pi}{\pi}$ decays.
 In addition, it can be seen from Eq.(\ref{hardblock})
 that the spectator scattering corrections depend
 strongly on the inverse moment parameter
 ${\lambda}_{B}$ given in Eq.(\ref{lamdef}).
 Recently, the value of ${\lambda}_{B}$ is an
 increasing concern of theoretical and
 experimental physicists \cite{Beneke5,Beneke4,Braun,BaBarBA1,BaBarBA2,lambda}.
 A scrutiny of parameter ${\lambda}_{B}$ becomes
 imperative.
 In this paper, we will give some information
 on ${\lambda}_{B}$ required by present experimental
 data of $B_{u,d}$ ${\to}$ $K \bar{K}$, ${\pi}K$,
 ${\pi}{\pi}$ decays.

 \section{numerical analysis and discussions}
 \label{sec03}
 With the conventions in Ref. \cite{Beneke2},
 the decay amplitudes for $B_{u,d}$ ${\to}$ $\pi K$,
 $K \bar{K}$, $\pi \pi $ decays within the QCDF
 framework can be written as
  \begin{eqnarray}
 {\cal A}_{ B^- \to \pi^- \bar{K}^0 }
  &=&
 \sum\limits_{p=u,c} V_{pb}V_{ps}^{\ast}
 A_{ \pi K }
 \Big\{ \alpha_{4}^{p} - \frac{1}{2} \alpha_{4,{\rm EW}}^{p}
 + {\delta}_{pu} \beta_{2} + \beta_{3}^{p} +
 \beta_{3,{\rm EW}}^{p} \Big\}
 \label{bm2pimkz}, \\
 \sqrt{2} {\cal A}_{ B^- \to \pi^0 K^- }
 &=&
 \sum\limits_{p=u,c} V_{pb}V_{ps}^{\ast}
 \Big\{ A_{ \pi K} \Big[ \delta_{pu} ( \alpha_1 + \beta_2 )
 + \alpha_4^p + \alpha_{4,{\rm EW}}^p + \beta_3^p
 + \beta_{3,{\rm EW}}^p \Big]
 \nonumber \\
 & & + A_{ K \pi } \Big[ \delta_{pu} \alpha_2
 + \frac{3}{2} \alpha_{3,{\rm EW}}^p \Big] \Big\}
 \label{amp2}, \\
 {\cal A}_{ \bar{B}^0 \to \pi^+ K^- }
 &=& \sum\limits_{p=u,c} V_{pb}V_{ps}^{\ast}
  A_{ \pi K} \Big\{ \delta_{pu} \alpha_1
 + \alpha_4^p + \alpha_{4,{\rm EW}}^p
 + \beta_3^p - \frac{1}{2} \beta_{3,{\rm EW}}^p
   \Big\}
 \label{amp3}, \\
 \sqrt{2} {\cal A}_{ \bar{B}^0 \to \pi^0 \bar{K}^0 }
 &=&
 \sum\limits_{p=u,c} V_{pb}V_{ps}^{\ast}
 \Big\{ A_{ \pi K} \Big[ - \alpha_4^p
 + \frac{1}{2} \alpha_{4,{\rm EW}}^p - \beta_3^p
 + \frac{1}{2} \beta_{3,{\rm EW}}^p \Big]
 \nonumber \\
 & & + A_{ K \pi } \Big[ \delta_{pu} \alpha_2
 + \frac{3}{2} \alpha_{3,{\rm EW}}^p \Big] \Big\}
 \label{b02pi0k0}, \\
 {\cal A}_{ B^- \to K^0 \bar{K}^0 }
 &=& \sum\limits_{p=u,c} V_{pb}V_{pd}^{\ast}
 A_{ K K} \Big\{ \alpha_4^p
 - \frac{1}{2} \alpha_{4,{\rm EW}}^p
 + \delta_{pu} \beta_2 + \beta_{3}^{p}
 + \beta_{3,{\rm EW}}^p \Big\}
 \label{bm2kk}, \\
 {\cal A}_{ \bar{B}^0 \to K^- K^+}
  &=& \sum\limits_{p=u,c} V_{pb}V_{pd}^{\ast}
  \Big\{ B_{ \bar{K} K} \Big[ \delta_{pu} b_1 + b_4^p
  + b_{4,{\rm EW}}^p \Big] + B_{ K \bar{K} } \Big[
  b_4^p - \frac{1}{2} b_{4,{\rm EW}}^p \Big] \Big\}
  \label{amp4}, \\
 {\cal A}_{ \bar{B}^0 \to \bar{K}^0 K^0}
  &=& \sum\limits_{p=u,c} V_{pb}V_{pd}^{\ast}
  \Big\{ A_{ \bar{K} K} \Big[ \alpha_4^p
  - \frac{1}{2} \alpha_{4,{\rm EW}}^p + \beta_3^p
  + \beta_4^p - \frac{1}{2} \beta_{3,{\rm EW}}^p
  - \frac{1}{2} \beta_{4,{\rm EW}}^p \Big]
  \nonumber \\ & &
  + B_{ K \bar{K} } \Big[
  b_4^p - \frac{1}{2} b_{4,{\rm EW}}^p \Big] \Big\}
  \label{b02kzkz}, \\
 \sqrt{2} {\cal A}_{ B^- \to \pi^- \pi^0 }
  &=& \sum\limits_{p=u,c} V_{pb}V_{pd}^{\ast}
  A_{ \pi \pi }
  \Big\{ \delta_{pu} ( \alpha_1 + \alpha_2 )
  + \frac{3}{2} ( \alpha_{3,{\rm EW}}^p
  + \alpha_{4,{\rm EW}}^p) \Big\}
  \label{amp5}, \\
 {\cal A}_{ \bar{B}^0 \to \pi^+ \pi^- }
  &=& \sum\limits_{p=u,c} V_{pb}V_{pd}^{\ast}
  A_{ \pi \pi }
  \Big\{ \delta_{pu} ( \alpha_1 + \beta_1 )
  + \alpha_{4}^p + \alpha_{4,{\rm EW}}^p
  + \beta_3^p + 2 \beta_4^p
  \nonumber \\ & &
  - \frac{1}{2} \beta_{3,{\rm EW}}^p
  + \frac{1}{2} \beta_{4,{\rm EW}}^p \Big\}
  \label{amp6}, \\
 -{\cal A}_{ \bar{B}^0 \to \pi^0 \pi^0 }
  &=& \sum\limits_{p=u,c} V_{pb}V_{pd}^{\ast}
  A_{ \pi \pi } \Big\{ \delta_{pu} ( \alpha_2 - \beta_1 )
  - \alpha_{4}^p + \frac{3}{2}\alpha_{3,{\rm EW}}^p
  + \frac{1}{2} \alpha_{4,{\rm  EW}}^p
  \nonumber \\ & &
  - \beta_3^p -2 \beta_4^p
  + \frac{1}{2} \beta_{3,{\rm EW}}^p
  - \frac{1}{2} \beta_{4,{\rm EW}}^p) \Big\}
  \label{amp7}.
  \end{eqnarray}

 For the sake for convenient discussion, we reiterate the
 expressions of the annihilation coefficients \cite{Beneke2},
 \begin{eqnarray}
  {\beta}_{i}^{p} &=& b_{i}^{p} B_{M_{1}M_{2}}/A_{M_{1}M_{2}}
   \label{betai}, \\
   b_{1} &=&
   \frac{C_{F}}{N_{c}^{2}}\, C_{1} A_{1}^{i},
   \quad \quad \quad
   b_{2} =
   \frac{C_{F}}{N_{c}^{2}}\, C_{2} A_{1}^{i}
   \label{b12}, \\
   b_{3}^{p} &=&
   \frac{C_{F}}{N_{c}^{2}}\,
   \Big[ C_{3} A_{1}^{i} + C_{5}( A_{3}^{i} + A_{3}^{f} )
   +N_{c} C_{6} A_{3}^{f} \Big]
   \label{b3}, \\
   b_{4}^{p} &=&
   \frac{C_{F}}{N_{c}^{2}}\,
   \Big[ C_{4} A_{1}^{i} + C_6 A_2^i \Big]
   \label{b4}, \\
   b_{3,\rm EW}^p &=&
   \frac{C_{F}}{N_{c}^{2}}\,
   \Big[ C_9 A_1^i + C_7 ( A_3^i + A_3^f )
   + N_c C_8 A_3^f \Big]
   \label{b3ew}, \\
   b_{4,\rm EW}^p &=&
   \frac{C_{F}}{N_{c}^{2}}\,
   \Big[ C_{10} A_1^i + C_8 A_2^i \Big]
   \label{b4ew}.
   \end{eqnarray}

 Numerically, coefficients of $b_{3,\rm EW}^p$ and
 $b_{4,\rm EW}^p$ are negligible compared with the
 other effective coefficients due to the small
 electroweak Wilson coefficients, and so their
 effects would be not discussed in this paper.

 In order to illustrate the contributions of
 annihilation and spectator scattering, we
 explore three parameter scenarios in which
 certain parameters are changed freely.
 \begin{itemize}
 \item Scenario I:
 $B_{u,d}$ ${\to}$ $\pi K$ and $K \bar{K}$ decays, including
 the $\pi K$ puzzle and pure annihilation decay
 $B_{d}$ $\to$ $K^- K^+$, are studied in detail.
 Combining the latest experimental data on the
 $CP$-averaged branching ratios, direct and
 mixing-induced $CP$-asymmetries,
 total 14 observables (see Table.\ref{pikbr},
 \ref{pikdcp}, \ref{pikmcp}) for seven
 $B_{u,d}$ ${\to}$ $\pi K$, $K \bar{K}$ decay modes
 [see Eq.(\ref{bm2pimkz}---\ref{b02kzkz})],
 the fit on four parameters ($\rho^f_A$, $\phi^f_A$)
 and ($\rho^i_A$, $\phi^i_A$) is performed with
 the fixed value $\lambda_B$ $=$ 0.2 GeV and
 the approximation ($\rho_H$, $\phi_H$) =
 ($\rho^i_A$, $\phi^i_A$),
 where ($\rho^f_A$, $\phi^f_A$), ($\rho^i_A$, $\phi^i_A$)
 and ($\rho_H$, $\phi_H$) are assumed to be universal for
 factorizable annihilation amplitudes,
 nonfactorizable annihilation amplitudes
 and spectator scattering corrections,
 respectively.
 \item Scenario II: $B_{u,d}$ ${\to}$ $\pi K$, $K \bar{K}$
 and $\pi \pi$ decays, including $\pi \pi$ puzzle,
 are studied.
 Combining the latest experimental data on the
 $CP$-averaged branching ratios, direct and
 mixing-induced $CP$-asymmetries,
 total 21 observables (see Table.\ref{pikbr},
 \ref{pikdcp}, \ref{pikmcp}) for ten $B_{u,d}$
 ${\to}$ $\pi K$, $K \bar{K}$, $\pi \pi$ decay modes
 [see Eq.(\ref{bm2pimkz}---\ref{amp7})],
 the fit on five parameters
 ($\rho^f_A$, $\phi^f_A$), ($\rho^i_A$, $\phi^i_A$)
 and $\lambda_B$ is performed with the approximation
 ($\rho_H$, $\phi_H$) = ($\rho^i_A$, $\phi^i_A$). 
 \item 
 Scenario III: As a general scenario, to clarify the relative 
 strength among ($\rho^f_A$, $\phi^f_A$), ($\rho^i_A$, $\phi^i_A$) 
 and ($\rho_H$, $\phi_H$), and check whether the approximation
 ($\rho_H$, $\phi_H$) = ($\rho^i_A$, $\phi^i_A$) is allowed or not,
 a fit on such six free parameters is performed.
 \end{itemize}

 Other input parameters used in our evaluation
 are summarized in Appendix \ref{app02}.
 Our fit approach is illustrated in detail
 in Appendix \ref{app03}.

 \subsection{Scenario I}
 \label{sec0301}
\begin{figure}[htb]
 \begin{center}
 \subfigure[]{\includegraphics[width=0.4\textwidth]{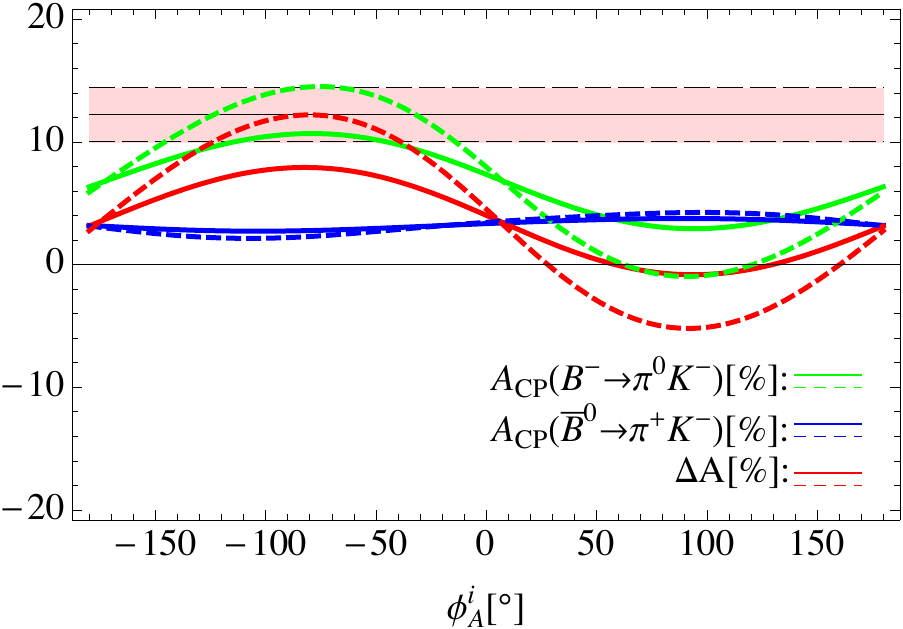}}
 \quad
 \subfigure[]{\includegraphics[width=0.4\textwidth]{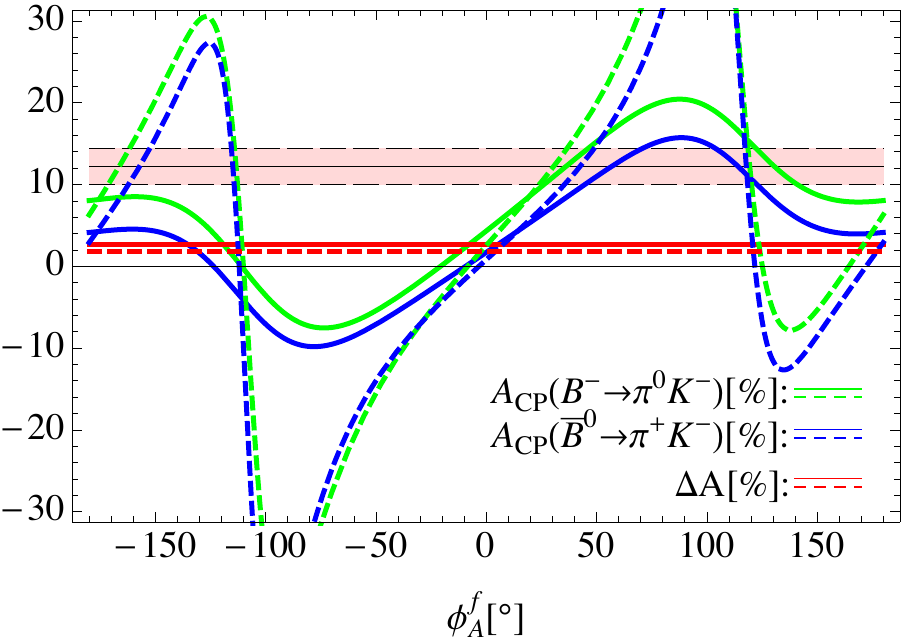}}
 \caption{The direct $CP$ asymmetries $A_{CP}(B^- \to \pi^0 K^-)$,
 $A_{CP}(\bar{B}^0 \to \pi^+ K^-)$ and their difference
 $\Delta A$ via (a) parameters ($\rho_A^i$, $\phi_A^{i}$)
 with  $\rho_A^{f}$ $=$ $\phi_A^{f}$ $=$ $0$; and
 (b) parameters ($\rho_A^f$, $\phi_A^f$) with $\rho_A^{i}$ $=$
 $\phi_A^{i}$ $=$ $0$, where the solid and dashed lines
 correspond to $\rho_A^{i,f}$ $=$ $1$ and $2$, respectively;
 The shaded band is the experimental result for $\Delta A$
 with $1 \sigma$ error.}
 \label{cpanni}
 \end{center}
 \end{figure}

 Comparing Eq.(\ref{amp2}) with Eq.(\ref{amp3}),
 it can be clearly seen that
 $\sqrt{2} {\cal A}_{ B^- \to \pi^0 K^-}$ $\simeq$
 ${\cal A}_{ \bar{B}^0 \to \pi^+ K^-}$
 if $\delta_{pu} \alpha_2$ $+$
 $\frac{3}{2} \alpha_{3,{\rm EW}}^p$
 is negligible compared with
 $\delta_{pu} \alpha_1$ $+$ $\alpha_4^p$.
 Hence it is expected $\Delta A$ $\simeq$ 0 in SM,
 which significantly disagrees with the current
 experimental data in Eq.(\ref{acppi}), this is
 the so-called $\pi K$ puzzle.
 To resolve the $\pi K$ puzzle, one possible solution
 is that there is a large complex contributions from
 $\delta_{pu} \alpha_2$ $+$
 $\frac{3}{2} \alpha_{3,{\rm EW}}^p$.
 Many proposals have been offered, such as the
 enhancement of color-suppressed
 tree amplitude $\alpha_2$ in Ref.\cite{Cheng1},
 significant new physics corrections to the
 electroweak penguin coefficient $\alpha_{3,{\rm EW}}^p$
 in Ref.\cite{pipipuz}, and so on.
 Indeed, it has been shown \cite{Beneke2} that the
 coefficients $\alpha_2$ and $\alpha_{3,{\rm EW}}^p$
 are seriously affected by spectator scattering
 corrections within QCDF framework.
 Consequently, the nonfactorizable spectator scattering
 parameters $X_H$ or ($\rho_H$, $\phi_H$) will have
 great influence on the observable $\Delta A$.
 Furthermore, a scrutiny of difference between Eq.(\ref{amp2})
 and Eq.(\ref{amp3}), another possible resolution to
 the $\pi K$ puzzle might be provided by annihilation
 contributions, such as coefficient $\beta_2$,
 as suggested in Ref.\cite{zhu2}.
 If so, then $\Delta A$ will depend strongly on
 the nonfactorizable annihilation parameters
 ($\rho_A^i$, $\phi_A^i$) because $\beta_2$ is
 proportional to $A_1^i$ in Eq.(\ref{b12}).
 Additionally, it can be seen from Eq.(\ref{amp2}) and
 Eq.(\ref{amp3}) that annihilation coefficient $\beta_3^p$
 contributes
 to amplitudes both ${\cal A}_{ B^- \to \pi^0 K^-}$
 and ${\cal A}_{ \bar{B}^0 \to \pi^+ K^-}$.
 If $\beta_3^p$ could offer a large strong phase,
 then its effect should contribute
 to the direct $CP$ asymmetries $A_{CP}(B^- \to \pi^0 K^-)$
 and $A_{CP}(\bar{B}^0 \to \pi^+ K^-)$ rather
 than $\Delta A$.
 Due to the fact that the lion's share of $\beta_3^p$
 comes from $N_{c} C_6 A_3^f$ in Eq.(\ref{b3}),
 the direct $CP$ asymmetries
 $A_{CP}(B^- \to \pi^0 K^-)$ and
 $A_{CP}(\bar{B}^0 \to \pi^+ K^-)$
 should vary greatly with the factorizable annihilation
 parameters $X_A^f$, while $\Delta A$ should be
 insensitive to variation of parameters ($\rho_A^f$, $\phi_A^f$).
 The above analysis and speculations are
 confirmed by Fig.\ref{cpanni}.

\begin{figure}[htb]
 \begin{center}
 \includegraphics[width=0.4\textwidth]{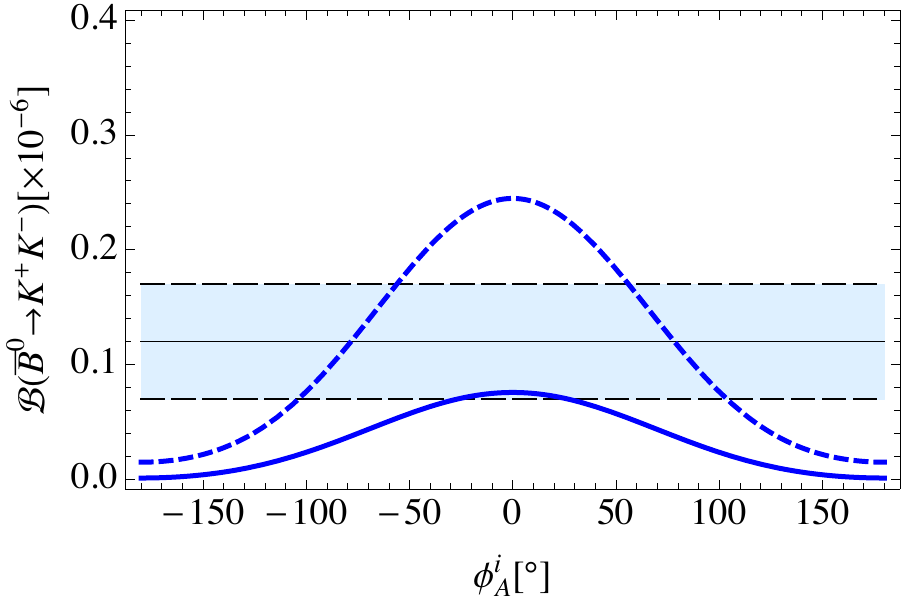}
 \caption{The dependence of branching ratio
  ${\cal B}(\bar{B}^0 \to K^- K^+)$
  on nonfactorizable annihilation parameters
  ($\rho_A^i$, $\phi_A^{i}$).
  The notes are the same as Fig.\ref{cpanni}.}
 \label{branni}
 \end{center}
 \end{figure}

 From Eq.(\ref{amp4}), it is seen that the amplitude
 ${\cal A}_{ \bar{B}^0 \to K^- K^+}$
 depends heavily on coefficients $\beta_1$ and $\beta_4^p$,
 which are closely associated with the nonfactorizable
 annihilation parameter $X_A^i$ only.
 The factorizable annihilation contributions vanish due
 to the isospin symmetry, which is consistent with the
 pQCD calculation \cite{xiao1}.
 The large branching ratio Eq.(\ref{HFAGKK}) would
 appeal for large nonfactorizable
 annihilation parameter $X_A^i$ or $\rho_A^i$.
 The dependence of branching ratio
 ${\cal B}(\bar{B}^0 \to K^- K^+)$
 on the parameters ($\rho_A^i$, $\phi_A^i$)
 is displayed in Fig.\ref{branni}.

 \begin{figure}[htb]
 \begin{center}
 \subfigure[]{\includegraphics[width=0.4\textwidth]{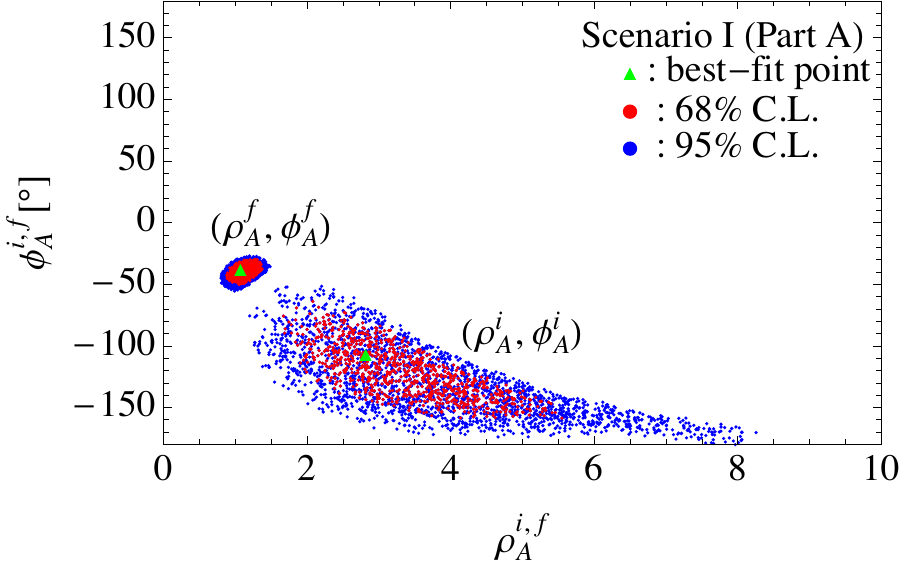}}
 \quad
 \subfigure[]{\includegraphics[width=0.4\textwidth]{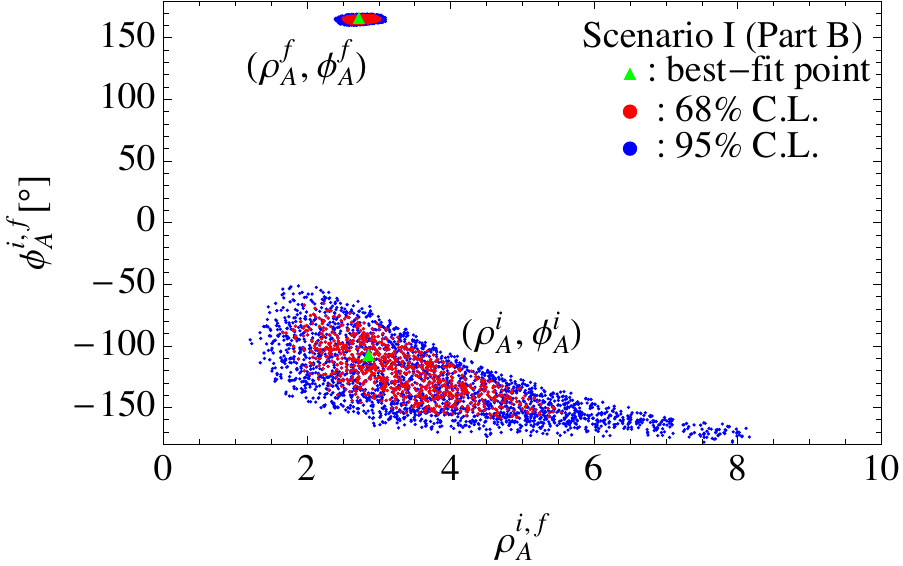}}
 \caption{The allowed regions of annihilation parameters at
 $68\%$ C. L. and $95\%$ C. L. in $(\rho_A^{i,f},\phi_A^{i,f})$
 planes, where the best-fit points of part A and B correspond
 to $\chi^2_{\rm min}$ $=$ $2.47$ and $\chi^2_{\rm min}$ $=$ $2.46$,
 respectively.}
 \label{ParaSpacI}
 \end{center}
 \end{figure}
 \begin{table}[htb]
 \caption{Numerical results of annihilation parameters in scenario I.}
 \label{pikfit}
 \begin{ruledtabular}
 \begin{tabular}{lcccc}
         & $\rho_H$ $=$ $\rho_A^i$
         & $\phi_H$ $=$ $\phi_A^i\,[^{\circ}]$
         & $\rho_A^f$ & $\phi_A^f[^{\circ}]$ \\ \hline
  Part A & $2.82^{+2.73}_{-1.15}$ & $-108^{+44}_{-50}$
         & $1.07^{+0.30}_{-0.20}$ & $-40^{+10}_{-11}$    \\
  Part B & $2.86^{+2.68}_{-1.20}$ & $-108^{+42}_{-51}$
         & $2.72^{+0.30}_{-0.22}$ & $166^{+3}_{-4}$
 \end{tabular}
 \end{ruledtabular}
 \end{table}
 \begin{table}[htb]
 \caption{The CP-averaged branching ratios (in units of $10^{-6}$)
 of $B$ ${\to}$ $\pi K$, $K \bar{K}$, $\pi \pi$ decays.
 For the Part A results of scenario I and II,
 the first and second theoretical uncertainties are
 caused by the CKM and other input parameters, respectively.}
 \label{pikbr}
 \begin{ruledtabular}
 \begin{tabular}{lcccc}
 \multicolumn{1}{c}{Decay Mode} & Exp. \cite{HFAG} & scenario I
 & scenario II  & S4 \cite{Beneke2} \\ \hline
   $B^- \to \pi^- \bar{K}^0$
 & $23.79 \pm 0.75$
 & $20.53^{+1.52+4.28}_{-0.65-3.87}$
 & $21.54^{+1.60+4.40}_{-0.68-3.99}$
 & $20.3$
   \\
   $B^- \to \pi^0 K^-$
 & $12.94^{+0.52}_{-0.51}$
 & $11.29^{+0.88+2.14}_{-0.45-1.96}$
 & $11.78^{+0.92+2.20}_{-0.47-2.01}$
 & $11.7$
   \\
   $\bar{B}^0 \to \pi^+ K^-$
 & $19.57^{+0.53}_{-0.52}$
 & $17.54^{+1.34+3.61}_{-0.65-3.27}$
 & $18.51^{+1.41+3.73}_{-0.67-3.38}$
 & $18.4$
   \\
   $\bar{B}^0 \to \pi^0 \bar{K}^0$
 & $9.93 \pm 0.49$
 & $8.05^{+0.60+1.84}_{-0.27-1.65}$
 & $8.60^{+0.65+1.90}_{-0.29-1.72}$
 & $8.0$
   \\ \hline
   $B^- \to K^- K^0$
 & $1.19 \pm 0.18$
 & $1.45^{+0.13+0.32}_{-0.09-0.29}$
 & $1.51^{+0.13+0.32}_{-0.09-0.29}$
 & $1.46$
   \\
   $\bar{B}^0 \to K^- K^+$
 & $0.12 \pm 0.05$
 & $0.13^{+0.01+0.02}_{-0.01-0.02}$
 & $0.15^{+0.02+0.02}_{-0.01-0.02}$
 & $0.07$
   \\
   $\bar{B}^0 \to K^0 \bar{K}^0$
 & $1.21 \pm 0.16$
 & $1.22^{+0.11+0.27}_{-0.08-0.24}$
 & $1.32^{+0.12+0.27}_{-0.08-0.25}$
 & $1.58$
   \\ \hline
   $B^- \to \pi^- \pi^0$
 & $5.48^{+0.35}_{-0.34}$
 & $5.20^{+0.64+1.11}_{-0.47-1.00}$
 & $5.59^{+0.68+1.15}_{-0.51-1.04}$
 & $5.1$
   \\
   $\bar{B}^0 \to \pi^+ \pi^-$
 & $5.10 \pm 0.19$
 & $5.88^{+0.66+1.66}_{-0.49-1.45}$
 & $5.74^{+0.64+1.63}_{-0.47-1.42}$
 & $5.2$
   \\
   $\bar{B}^0 \to \pi^0 \pi^0$
 & $1.91^{+0.22}_{-0.23}$
 & $1.67^{+0.22+0.25}_{-0.19-0.23}$
 & $2.13^{+0.29+0.32}_{-0.24-0.29}$
 & $0.7$
   \\ \hline
   $R_{+-}^{\pi \pi}$
 & $1.99 \pm 0.15$
 & $1.64^{+0.06+0.13}_{-0.06-0.11}$
 & $1.80^{+0.07+0.17}_{-0.07-0.13}$
 & $1.82$
   \\
   $R_{00}^{\pi \pi}$
 & $0.75 \pm 0.09$
 & $0.57^{+0.06+0.16}_{-0.06-0.12}$
 & $0.74^{+0.08+0.22}_{-0.08-0.17}$
 & $0.27$
   \end{tabular}
   \end{ruledtabular}
   \end{table}
  \begin{table}[htb]
  \caption{The direct CP asymmetries (in units of $10^{-2}$)
   of $B$ ${\to}$ $\pi K$, $K \bar{K}$, $\pi \pi$ decays.
   The notes on uncertainties are the same as Table\ref{pikbr}.}
  \label{pikdcp}
  \begin{ruledtabular}
  \begin{tabular}{lcccc}
  \multicolumn{1}{c}{Decay Mode} & Exp. \cite{HFAG} & scenario I
  & scenario II  & S4 \cite{Beneke2} \\ \hline
    $B^- \to \pi^- \bar{K}^0$
  & $-1.5 \pm 1.9$
  & $-0.05^{+0.00+0.13}_{-0.00-0.15}$
  & $-0.17^{+0.01+0.14}_{-0.01-0.15}$
  & $0.3$
    \\
    $B^- \to \pi^0 K^-$
  & $4.0 \pm 2.1$
  & $3.2^{+0.2+0.6}_{-0.2-0.6}$
  & $2.5^{+0.1+0.6}_{-0.1-0.6}$
  & $-3.6$
    \\
    $\bar{B}^0 \to \pi^+ K^-$
  & $-8.2 \pm 0.6$
  & $-7.7^{+0.4+0.9}_{-0.4-0.9}$
  & $-9.1^{+0.4+0.9}_{-0.5-0.9}$
  & $-4.1$
    \\
    $\bar{B}^0 \to \pi^0 \bar{K}^0$
  & $-1 \pm 10$
  & $-10.3^{+0.6+0.9}_{-0.6-1.0}$
  & $-10.6^{+0.6+0.9}_{-0.6-0.9}$
  & $0.8$
    \\ \hline
    $\Delta A$
  & $12.2 \pm 2.2$
  & $10.9^{+0.6+0.9}_{-0.5-0.8}$
  & $11.6^{+0.6+0.9}_{-0.6-0.8}$
  & $0.5$
    \\ \hline
    $B^- \to K^- K^0$
  & $3.9 \pm 14.1$
  & $-0.6^{+0.0+3.2}_{-0.0-2.9}$
  & $2.0^{+0.1+3.4}_{-0.1-3.0}$
  & $-4.3$
    \\
    $\bar{B}^0 \to K^0 \bar{K}^0$
  & $-6 \pm 26$
  & $-17^{+1+2}_{-1-2}$
  & $-16^{+1+2}_{-1-2}$
  & $-11.5$
    \\ \hline
    $B^- \to \pi^- \pi^0$
  & $2.6 \pm 3.9$
  & $-1.1^{+0.1+0.1}_{-0.1-0.1}$
  & $-1.2^{+0.1+0.1}_{-0.1-0.1}$
  & $-0.02$
    \\
    $\bar{B}^0 \to \pi^+ \pi^-$
  & $29 \pm 5$
  & $19^{+1+4}_{-1-4}$
  & $24^{+2+5}_{-2-4}$
  & $10.3$
    \\
    $\bar{B}^0 \to \pi^0 \pi^0$
  & $43 \pm 24$
  & $46^{+3+6}_{+3-6}$
  & $38^{+2+6}_{-2-6}$
  & $-19.0$
  \end{tabular}
  \end{ruledtabular}
  \end{table}
  \begin{table}[htb]
  \caption{The mixing-induced $CP$ asymmetries (in units of $10^{-2}$)
   of $B$ ${\to}$ $\pi K$, $K \bar{K}$, $\pi \pi$ decays.
   The notes on uncertainties are the same as Table\ref{pikbr}.}
  \label{pikmcp}
  \begin{ruledtabular}
  \begin{tabular}{lcccc}
  \multicolumn{1}{c}{Decay Mode} & Exp. \cite{HFAG} & scenario I & scenario II  \\ \hline
   $\bar{B}^0 \to \pi^0 \bar{K}^0$
 & $57 \pm 17$
 & $78^{+3+1}_{-3-1}$
 & $79^{+3+1}_{-3-1}$
   \\ \hline
   $\bar{B}^0 \to K^- K^+$
 & ---
 & $-86^{+6+0}_{-5-0}$
 & $-86^{+6+0}_{-5-0}$
   \\
   $\bar{B}^0 \to K^0 \bar{K}^0$
 & $-108 \pm 49$
 & $-10^{+1+0}_{-1-0}$
 & $-11^{+1+0}_{-1-0}$
   \\ \hline
   $\bar{B}^0 \to \pi^+ \pi^-$
 & $-65 \pm 6$
 & $-59^{+11+2}_{-10-3}$
 & $-60^{+10+2}_{-10-2}$
   \\
   $\bar{B}^0 \to \pi^0 \pi^0$
 & ---
 & $77^{+6+1}_{-8-2}$
 & $77^{+7+1}_{-9-2}$
 \end{tabular}
 \end{ruledtabular}
 \end{table}

 To get more information on annihilation and spectator
 scattering, we perform a fit on the parameters $X_H$ $=$
 $X_A^i$ and $X_A^f$, considering the constraints of
 the $CP$-averaged branching ratios, direct and mixing-induced
 $CP$-asymmetries, from $B$ ${\to}$ $\pi K$, $K \bar{K}$ decays.
 The experimental data are summarized in
 the second column of Tables \ref{pikbr}-\ref{pikmcp}.
 Our fitting results are shown by
 Fig.\ref{ParaSpacI}, and the corresponding numerical
 results are listed in Table \ref{pikfit}-\ref{pikmcp}.

 It is found that two possible solutions entitled Part A
 and B in Table \ref{pikfit}, correspond to almost the
 same $(\rho_A^i, \phi_A^i)$ $\approx$
 $(2.8,-108^{\circ})$.
 The large errors on parameter $(\rho_A^i, \phi_A^i)$
 are mainly caused by the current loose experimental
 constraints on $CP$ asymmetries measurements for $B$
 ${\to}$ ${\pi}K$, $K \bar{K}$ decays.
 In principle, the pure annihilation
 $\bar{B}^0 \to K^- K^+$ decays whose amplitudes
 depend predominantly on $(\rho_A^i, \phi_A^i)$,
 besides the decays constants, should give rigorous
 constraint on $X_A^i$.
 It's a pity that the available measurement accuracy
 on its branching ratio is too poor to efficiently
 confine $(\rho_A^i, \phi_A^i)$ to some tiny spaces.
 The large $(\rho_A^i,\phi_A^i)$ mean large $X_A^i$
 and $X_H$, i.e.,
 there must exist large nonfactorizable
 annihilation and spectator scattering contributions to
 accommodate the current measurements.
 Our fit results on parameter $\rho_A^i$ provide a robust
 evidence to the educated guesswrok about $\rho_{Ad}^i$
 $=$ 2.5 in Ref.\cite{zhu2}.
 In fact, the strong phase $\phi_A^i$ educed from
 measurements of branching ratios for $B^0$ $\to$
 $K \bar{K}$ decays in Ref.\cite{zhu2} can have
 either positive or negative values with the magnitudes
 of $\gtrsim$ $100^{\circ}$ (see Fig.5 of Ref.\cite{zhu2}),
 where the positive value $\phi_A^i$ $=$ $+100^{\circ}$
 used in Ref.\cite{zhu2} will be excluded by our fit
 with much more experimental data on $B$ ${\to}$ $\pi K$,
 $K \bar{K}$ decays.
 The large value of $\phi_A^i$, corresponding to a large
 imaginary part of the enhanced complex corrections,
 also lends some support to the pQCD claim that the
 annihilation amplitudes can provide a large strong
 phase \cite{pqcd}.

 There are two possible solutions for the factorizable
 annihilation parameters, namely, Part A
 $(\rho_A^f,\phi_A^f)$ $\approx$ $(1.1,-40^{\circ})$
 and Part B
 $(\rho_A^f,\phi_A^f)$ $\approx$ $(2.7,166^{\circ})$.
 From Fig.\ref{ParaSpacI}, it can be seen that
 there is no overlap between the regions of
 $(\rho_A^f,\phi_A^f)$ and $(\rho_A^i,\phi_A^i)$
 at the 95\% confidence level, which indicates that
 it might be wrong to treat
 $(\rho_A^f,\phi_A^f)$ $=$ $(\rho_A^i,\phi_A^i)$ $=$
 $(\rho_A,\phi_A)$
 as universal parameters for nonfactorizable
 and factorizable annihilation topologies in
 pervious studies.
 Our fit results certify the suggestion of Ref.\cite{zhu1,zhu2}
 that different annihilation topologies should be
 parameterized by different annihilation parameters,
 i.e., $(\rho_A^f,\phi_A^f)$ $\neq$ $(\rho_A^i,\phi_A^i)$.
 Compared with the results of $(\rho_A^i,\phi_A^i)$,
 the errors on parameter $(\rho_A^f, \phi_A^f)$ are
 relatively small (see Table \ref{pikfit}), because
 the available measurements on branching ratios for
 $B$ ${\to}$ ${\pi}K$ decays are highly precise.
 The conjecture about $(\rho_A^f, \phi_A^f)$ in \cite{zhu2}
 is somewhat alike to our fit results of Part A.

 The value of term $(2X_A^f-X_A^f)$ in Eq.(\ref{af3})
 is about $(27.2-i26.2)$ with parameters for Part A and
 $(28.9-i25.5)$ for Part B, that is to say,
 these two solutions, Part A and B, will present
 similar factorizable annihilation contributions.
 Nevertheless, a small value of $\rho_A^f$ is more
 easily accepted by the QCDF approach \cite{Beneke2}.
 So with the best fit parameters of Part A in
 Table \ref{pikfit}, we present our evaluations
 on branching ratios, direct and mixing-induced
 $CP$ asymmetries for $B_{u,d}$ ${\to}$ $\pi K$,
 $K \bar{K}$, $\pi \pi$ decays
 in the ``scenario I'' column of Table \ref{pikbr},
 \ref{pikdcp} and \ref{pikmcp}, respectively.
 For comparison, the results of scenario S4 QCDF
 \cite{Beneke2} are also collected in the ``S4''
 column. It is easily found that all theoretical
 results are in good agreement with experimental
 data within errors. Especially, the difference
 $\Delta A$, which $\sim$ 0.5\% in scenario S4 QCDF,
 is enhanced to the experimental level $\sim$ 11\%.
 It is interesting that although $B$ $\to$ $\pi \pi$
 decays are not considered in the ``scenario I'' fit,
 all predictions on these decays, including the ratios
 $R_{+-}^{\pi\pi}$ and $R_{00}^{\pi\pi}$,
 are also in good consistence with the experimental
 measurements within errors, which implies that the
 $\pi K$ and $\pi \pi$ puzzles could be resolved
 by annihilation and spectator corrections,
 at the same time, without violating the agreement
 of other observables. The reason will be
 excavated in Scenario II.

 \subsection{Scenario II}
 \label{sec0302}
 From Eq.(\ref{amp5}), it is obviously found that the
 amplitude of $B^-\to\pi^-\pi^0$ decay is independent
 of annihilation contributions, and dominated by
 $\alpha_1$ $+$ $\alpha_2$.
 Moreover, comparing Eq.(\ref{amp6}) with Eq.(\ref{amp7}),
 it is easily found that the annihilation contributions
 are almost helpless for $R_{00}^{\pi \pi}$ puzzle due
 to ${\cal A}_{B^0 \to \pi^+ \pi^-}^{\rm anni}$
 $\simeq$ ${\cal A}_{B^0 \to \pi^0 \pi^0}^{\rm anni}$.
 So, the spectator scattering corrections, which play an
 important role in the color-suppressed coefficient
 $\alpha_2$ \cite{Beneke2,Cheng1,Cheng3}, would be another
 important key for the good results of scenario I,
 especially for $B$ $\to$ $\pi\pi$ decays.

 Within QCDF framework, besides $X_H$, the inverse
 moment $\lambda_B$ of $B$ wave function defined by
 Eq.(\ref{lamdef}) is another important quantity in
 evaluating the contributions of spectator scattering.
 Unfortunately, its value is hardly to be obtained
 reliably with theoretical methods until now, for instance
 $350{\pm}150$ MeV (200 MeV in scenario S2) in Ref.\cite{Beneke2},
 $200^{+250}_{-0}$ MeV in Ref.\cite{Beneke4}
 and $300{\pm}100$ MeV in Ref.\cite{Cheng1},
 though QCD sum rule prefer $460{\pm}110$ MeV
 at the scale of 1 GeV \cite{Braun}.
 Experimentally, the upper limit on parameter $\lambda_B$
 are set at the 90\% C.L. via measurements on branching
 fraction of radiative leptonic $B$ $\to$
 $\ell \bar{\nu}_{\ell} \gamma $ decay by BABAR collaboration,
 $\lambda_B$ $>$ 669 (591) MeV with different priors
 based on 232 million $B\bar{B}$ sample
 where the photon is not required to be sufficiently
 energetic in order not to sacrifice statistics \cite{BaBarBA1},
 and $\lambda_B$ $>$ 300 MeV based on 465 million
 $B\bar{B}$ pairs \cite{BaBarBA2}.
 Considering radiative and power corrections, an improved
 analysis is preformed in Ref.\cite{Beneke5} with the
 conclusion that present BABAR measurements cannot put
 significant constrains on $\lambda_B$ and that
 $\lambda_B$ $>$ 115 MeV from the experimental results
 \cite{BaBarBA2}.
 Anyway, the study of hadronic $B$ decays favors a
 relative small value of $\lambda_B$ $\approx$ 200 MeV
 to achieve a satisfactory description of color-suppressed
 tree decay modes \cite{lambda}.
 At the present time, the value of $\lambda_B$ is still
 a point of controversy.
 In the following analysis and evaluations, we treat
 $\lambda_B$ as a free parameter.
 \begin{figure}[htb]
 \begin{center}
 \includegraphics[width=0.3\textwidth]{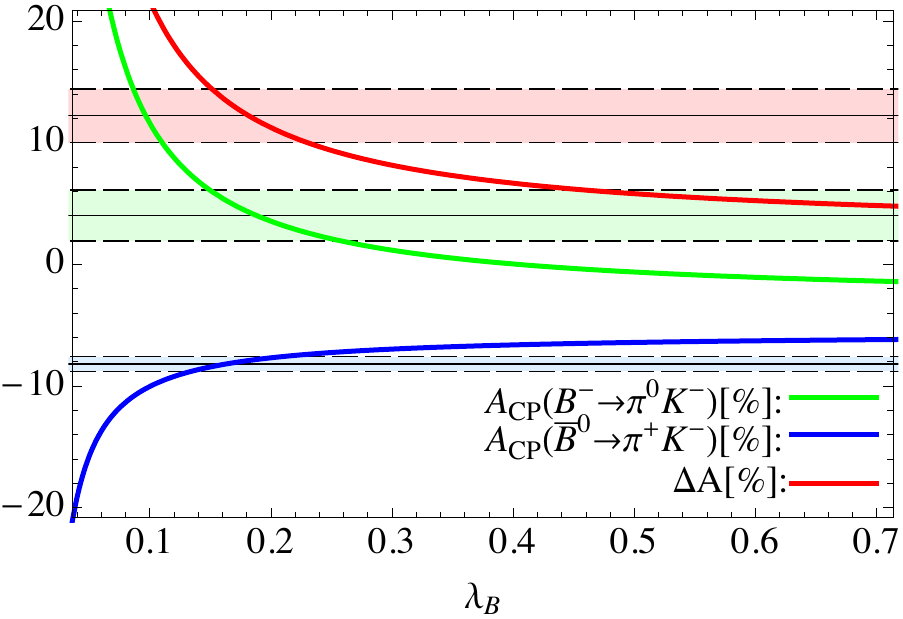}
 \caption{The dependance of the direct $CP$
  asymmetries $A_{CP}(B^- \to \pi^0 K^-)$,
  $A_{CP}(\bar{B}^0 \to \pi^+ K^-)$
  and their difference $\Delta A$ on $\lambda_B$
  (in unites of GeV) with the fitted annihilation
  parameters of scenario I (Part A).
  Their experimental results with $1\sigma$ error
  are shown by shaded bands with the same color
  as the lines.}
 \label{LBpik}
 \end{center}
 \end{figure}
 \begin{figure}[h]
 \begin{center}
 \subfigure[]{\includegraphics[width=0.3\textwidth]{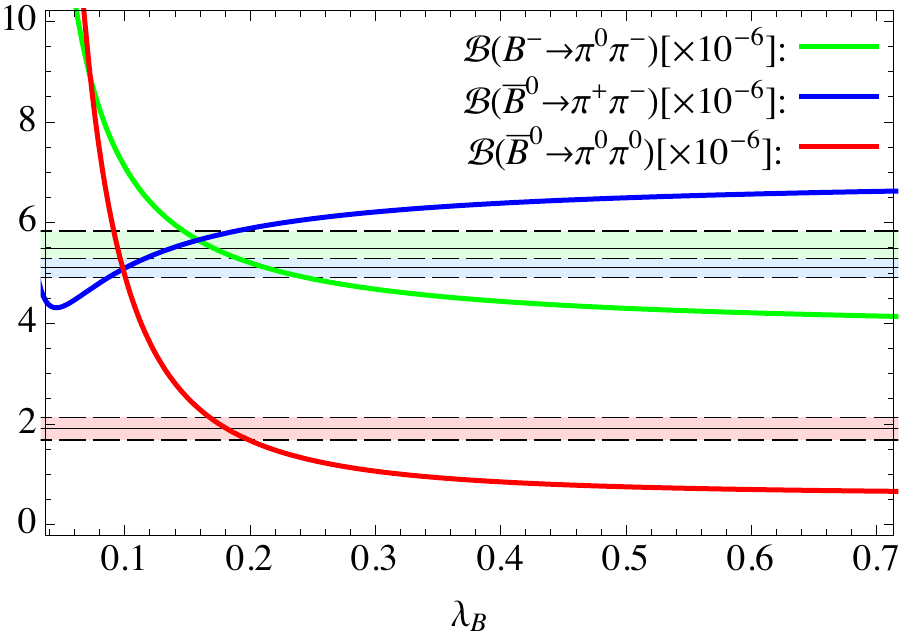}}
 \quad \quad
 \subfigure[]{\includegraphics[width=0.3\textwidth]{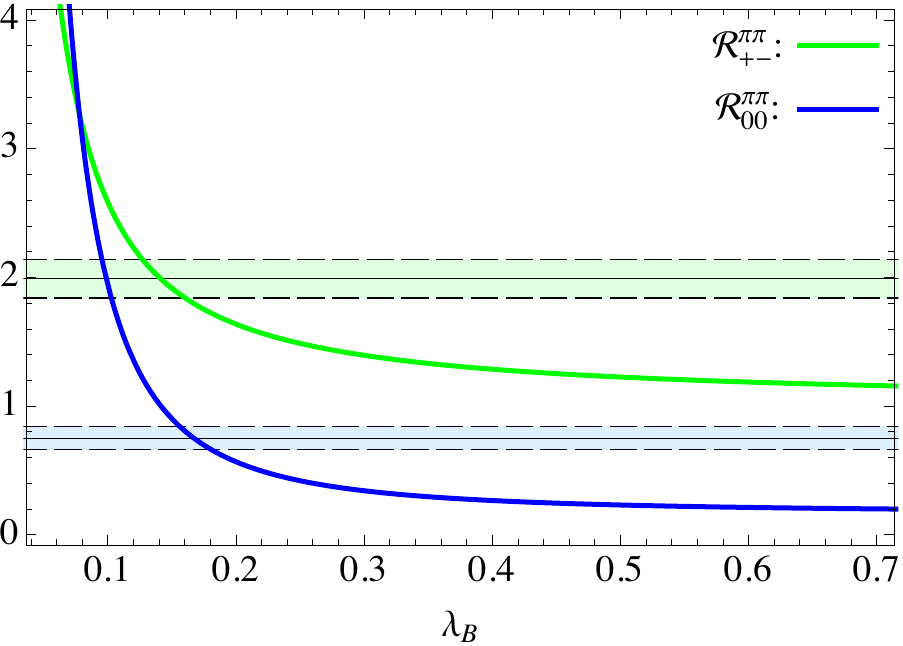}}
 \caption{The dependence of the branching fractions
  ${\cal B}(B^- \to \pi^- \pi^0 )$,
  ${\cal B}(\bar{B}^0 \to \pi^+ \pi^-)$,
  ${\cal B}(\bar{B}^0 \to \pi^0 \pi^0)$
  and ratios $R_{+-}^{\pi\pi}$,
  $R_{00}^{\pi\pi}$ on $\lambda_B$
  with the same notes as Fig.\ref{LBpik}.}
  \label{LBpipi}
  \end{center}
  \end{figure}
 \begin{figure}[h]
 \begin{center}
 \subfigure[]{\includegraphics[width=0.3\textwidth]{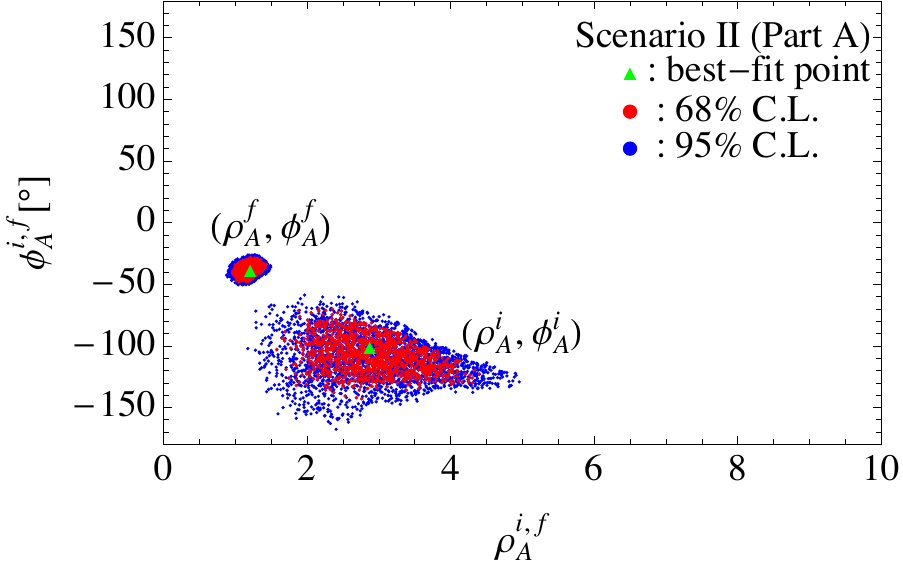}}
 \quad \quad
 \subfigure[]{\includegraphics[width=0.3\textwidth]{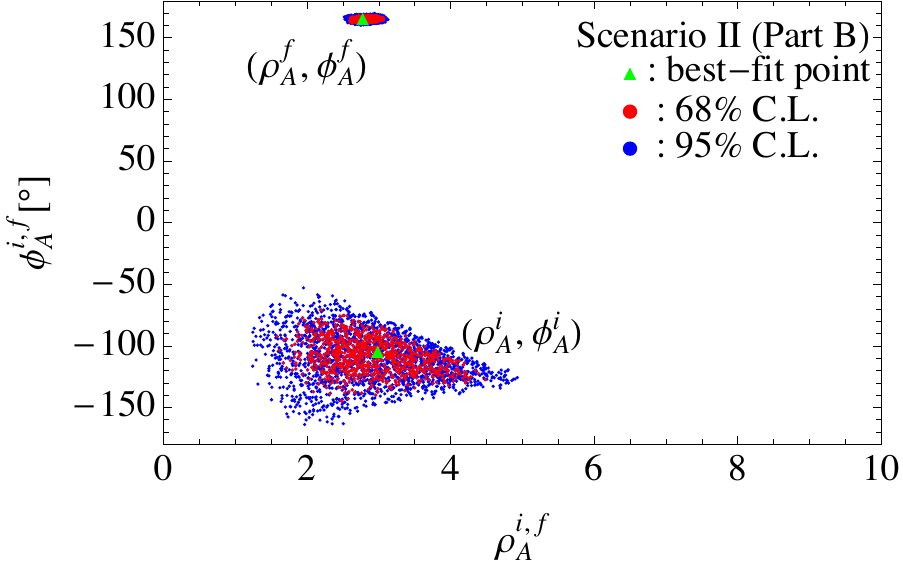}}
 \\
 \subfigure[]{\includegraphics[width=0.3\textwidth]{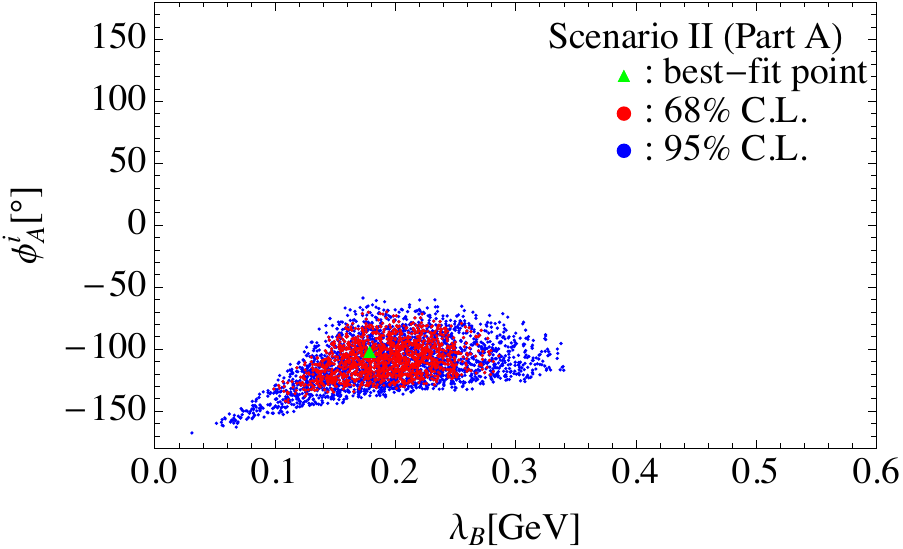}}
 \quad \quad
 \subfigure[]{\includegraphics[width=0.3\textwidth]{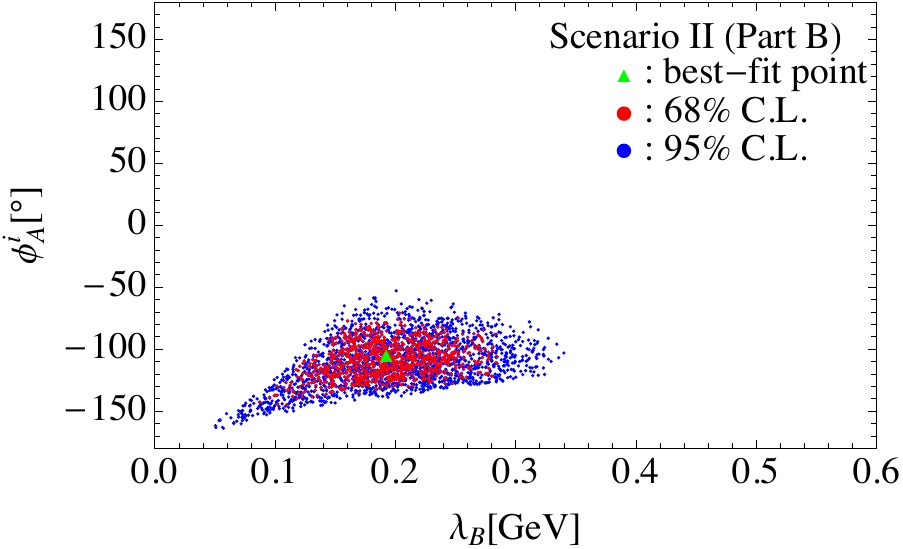}}
 \caption{The allowed regions of annihilation parameters
 ($\rho_A^{i,f}$, $\phi_A^{i,f}$) and $\lambda_B$ at
 $68\%$ C.L. and $95\%$ C.L.. The best-fit points
 of part A and B correspond to $\chi^2_{\rm min}=3.66$
 and $\chi^2_{\rm min}=3.67$, respectively.}
 \label{ParaSpacII}
 \end{center}
 \end{figure}
 \begin{table}[h]
 \caption{Numerical results of annihilation
  parameters and moment parameter $\lambda_B$
  in Scenario II.}
 \label{pipikfit}
 \begin{ruledtabular}
 \begin{tabular}{lccccc}
   & $\rho_A^i$ & $\phi_A^i[^{\circ}]$
   & $\rho_A^f$ & $\phi_A^f[^{\circ}]$
   & $\lambda_B$ [GeV] \\ \hline
   Part A
   & $2.88^{+1.52}_{-1.30}$ & $-103^{+33}_{-40}$
   & $1.21^{+0.22}_{-0.25}$ & $-40^{+12}_{-8}$
   & $0.18^{+0.11}_{-0.08}$
   \\
   Part B
   & $2.98^{+1.50}_{-1.40}$ & $-106^{+35}_{-39}$
   & $2.78^{+0.29}_{-0.18}$ & $165^{+4}_{-3}$
   & $0.19^{+0.09}_{-0.10}$
  \end{tabular}
  \end{ruledtabular}
  \end{table}

 To explicitly show the effects of
 spectator scattering contributions on $\pi K$ puzzle,
 dependance of $A_{CP}(B^- \to \pi^0 K^-)$,
 $A_{CP}(\bar{B}^0 \to \pi^+ K^-)$ and their
 difference $\Delta A$ on parameter $\lambda_B$
 are displayed in Fig.\ref{LBpik}.
 It is found that
 (1)
 observables of $A_{CP}(B^- \to \pi^0 K^-)$
 and $\Delta A$ are more sensitive to variation of
 $\lambda_B$ than $A_{CP}(\bar{B}^0 \to \pi^+ K^-)$
 in the region of $\lambda_B$ $\geq$ 100 MeV.
 The reason is aforementioned fact that coefficient
 $\alpha_2$ in amplitude ${\cal A}_{ B^- \to \pi^0 K^-}$
 [see Eq.(\ref{amp2})] receives significant
 spectator scattering corrections.
 A noticeable change of observables is
 easily seen in the low region of $\lambda_B$
 because spectator scattering corrections are
 inversely proportional to $\lambda_B$
 [see Eq.(\ref{lamdef}) and Eq.(\ref{hardblock})].
 (2)
 a relative small value of $\lambda_B$ $\in$
 [150 MeV, 220 MeV], as expected in \cite{lambda},
 is required to confront with available measurements.
 Especially, the value $\lambda_B$ $\approx$ 190 MeV
 provides a perfect description of the experimental
 data on $A_{CP}(B^- \to \pi^0 K^-)$,
 $A_{CP}(\bar{B}^0 \to \pi^+ K^-)$ and
 $\Delta A$ simultaneously.
 For $B$ $\to$ $\pi \pi$ decays,
 from Eqs.(\ref{amp5}-\ref{amp7}),
 it is easily seen that amplitude
 ${\cal A}_{ B^- \to \pi^- \pi^0 }$ $\propto$
 $\alpha_1$ + $\alpha_2$,
 ${\cal A}_{ \bar{B}^0 \to \pi^+ \pi^- }$
 $\propto$ $\alpha_1$,
 ${\cal A}_{ \bar{B}^0 \to \pi^0 \pi^0 }$
 $\propto$ $\alpha_2$.
 The coefficient $\alpha_2$, corresponding to the
 color-suppressed tree contribution, its value is
 small relative to $\alpha_1$, so the experimental
 data on $R_{+-}^{\pi \pi}$ can be well explained
 with scenario S4 QCDF where $X_A^i$ $=$ $X_A^f$
 and $\rho_A^{f,i}$ = 1 (see Table \ref{pikbr}).
 But as to observable $R_{00}^{\pi \pi}$ or/and
 branching ratio ${\cal B}(\bar{B}^0 \to \pi^0 \pi^0)$,
 an enhanced $\alpha_2$ is desirable. Hence,
 the nonfactorizable spectator scattering
 contributions, which have significant effects
 on $\alpha_2$, would play an important role in
 studying the color-suppressed tree $B$ decays,
 and possibly provide a solution to the $\pi \pi$ puzzle.
 The dependencies of the branching fractions of
 $B$ $\to$ $\pi\pi$ decays and ratios $R_{+-}^{\pi\pi}$,
 $R_{00}^{\pi\pi}$ on $\lambda_B$ are shown in
 Fig.\ref{LBpipi} where the fitted parameters of
 Part A in Table \ref{pikfit} is used.
 It is interesting that beside a large value $\rho_H$,
 a small value of $\lambda_B$ $\sim$ 200 MeV is also
 required to confront with experimental data on
 ${\cal B}(B \to \pi \pi)$, $R_{+-}^{\pi\pi}$ and
 $R_{00}^{\pi\pi}$.

 With the available experimental data on $B$ $\to$ $\pi \pi$,
 $\pi K$ and $K \bar{K}$ decays, we perform a comprehensive fit on
 both annihilation parameters ($\rho_{A}^{i,f}$, $\phi_{A}^{i,f}$)
 and $B$-meson wave function parameter $\lambda_B$.
 The allowed parameter spaces are shown in Fig.\ref{ParaSpacII},
 and the corresponding numerical results are summarized in
 Table \ref{pipikfit}.
 Like scenario I, there are two allowed spaces which are
 labelled by part A and B.
 It is easily found that (1) parameters $(\rho_A^i,\phi_A^i)$
 $=$ $(\rho_H,\phi_H)$ are still required to have large values
 (see Table \ref{pipikfit}), that is to say, it is necessary
 for penguin-dominated or color-suppressed tree $B$ decays to
 own large corrections from nonfactorizable annihilation and
 spectator scattering topologies.
 (2) There is still no overlap between the regions of
 $(\rho_A^f,\phi_A^f)$ and $(\rho_A^i,\phi_A^i)$ at the 95\%
 confidence level.
 (3) The cental values of $\rho_A^{i,f}$ are a little
 larger than those in scenario I. The uncertainties
 on $(\rho_A^i,\phi_A^i)$ are a little smaller than those
 in scenario I, because more processes from $B$ $\to$ $\pi \pi$
 decays are considered in fitting and the amplitudes for
 $B$ $\to$ $\pi \pi$ decays are sensitive to $X_A^i$ and $X_H$
 rather than $X_A^f$.
 (4) A small value of parameter $\lambda_B$ $\leq$ 350 MeV at
 the 95\% confidence level is strongly required to reconcile
 discrepancies between results of QCDF approach and available
 experimental data on $B$ $\to$ $\pi \pi$, $\pi K$ and
 $K \bar{K}$ decays.

 The two solutions of scenario II, Part A and B, will
 give similar results, as discussed before.
 With the best fit parameters of Part A in Table
 \ref{pipikfit}, we present our evaluations
 on branching ratios, direct and mixing-induced
 $CP$ asymmetries for $B_{u,d}$ ${\to}$ $\pi K$,
 $K \bar{K}$, $\pi \pi$ decays
 in the ``scenario II'' column of Table \ref{pikbr},
 \ref{pikdcp} and \ref{pikmcp}, respectively.
 It is found that
 the central values of branching ratios for
 $B$ $\to$ $\pi \pi$, $\pi K$ and $K \bar{K}$ decays,
 expect $\bar{B}^0$ $\to$ $\pi^+ \pi^-$ decay,
 with the Part A parameters of scenario II,
 are a little larger than those of scenario I
 (see Table \ref{pikbr}),
 because a bit larger values of $\rho_A^{i,f}$
 and a bit smaller value of $\lambda_B$ than those of
 scenario I are taken in scenario II.
 Compared with results of scenario S4 QCDF, agreement
 between theoretical results within two scenarios and
 experimental measurements is improved, especially for
 the observables $\Delta A$, $R_{00}^{\pi \pi}$
 and $A_{CP}(B^0 \to \pi \pi)$.
 \subsection{Scenario III}
 \label{sec0303}
 
 The above analyses and results are based on the assumption that 
 $X_A^{i}$ $=$ $X_{H}$ (i.e. $(\rho_A^{i}, \phi_A^{i})$ $=$ 
 $(\rho_H, \phi_H)$) for simplicity.
 While, there is no compellent requirement for 
 such simplification, except for the fact that 
 wave functions of $B$ mesons are involved in the convolution
 integrals of both spectator scattering and nonfactorizable annihilation
 corrections, but are irrelevant to the factorable annihilation amplitudes.
 So, as a general scenario~(named scenario III), we would reevaluate the 
 strength of annihilation and hard-spectator contributions without any simplification for the parameters 
 $(\rho_A^{i}, \phi_A^{i})$, $(\rho_A^{i}, \phi_A^{i})$ and $(\rho_H, \phi_H)$.
 
 \begin{figure}[t]
 \begin{center}
\includegraphics[width=0.35\textwidth]{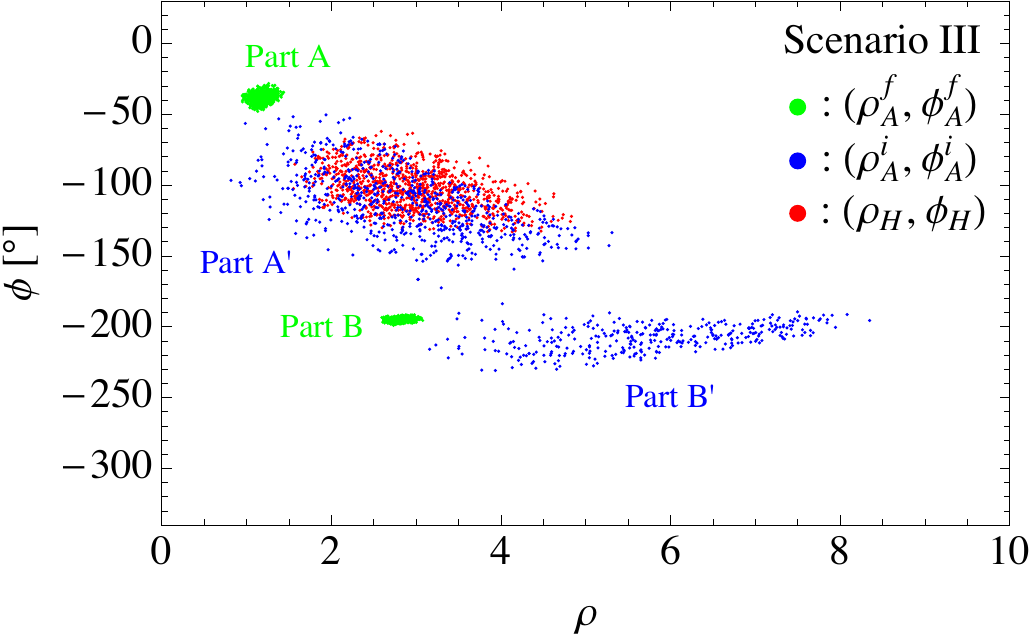}
 \caption{The allowed regions of annihilation and hard-spectator parameters
 ($\rho_A^{f}$, $\phi_A^{f}$), ($\rho_A^{i}$, $\phi_A^{i}$) and ($\rho_H$, $\phi_H$) at
 $68\%$ C.L.. The two solutions of  ($\rho_A^{f}$, $\phi_A^{f}$) and 
 ($\rho_A^{i}$, $\phi_A^{i}$) are labeled as Part A, B and 
 $\rm A^{\prime}$, $\rm B^{\prime}$, respectively.}
  \label{SpecFig}
  \end{center}
  \end{figure}

 Considering the constraints from observables of $B_{u,d}$ ${\to}$ $K \bar{K}$, ${\pi}K$ and ${\pi}{\pi}$ 
 decays, a fit for the annihilation and hard-spectator parameters is performed again. 
 In this fit, $(\rho_A^{f}, \phi_A^{f})$, $(\rho_A^{i}, \phi_A^{i})$ and $(\rho_H, \phi_H)$ are 
 treated as six free parameters.
 Moreover, from the hard-spectator corrections illustrated by Eq. (\ref{hardblock}), 
 it can be seen that $\lambda_B$ and $X_H$ are always combined together. 
 
 Although the inverse moment $\lambda_B$ of $B$ wave function could be
 determined or constricted by further experiments \cite{Beneke5,BaBarBA1,BaBarBA2,lambda},
 $\lambda_B$ is more like a free parameter for the moment due to loose limitation
 on it. So it is impossible to strictly bound on $\lambda_B$ and $X_H$ simultaneously
 due to the interference effects between them. 
 In our following fit, we will fix $\lambda_B=200\,{\rm MeV}$.
 Our fitting results at $68\%$ C.L. are presented in Fig.~\ref{SpecFig},  
 where the range of ${\phi}$ ${\in}$ $[-360^{\circ},0^{\circ}]$ is
 assigned to illustrate their relative magnitude.
 Numerically, we get
 \begin{eqnarray}
 &&(\rho_A^{f}, \phi_A^{f}[^{\circ}])=\left\{ \begin{array}{ll}
 & (1.18^{+0.26}_{-0.23}, -40^{+12}_{-8})\qquad \text{Part A } \\
& (2.79^{+0.26}_{-0.20}, -196^{+5}_{-3})\qquad  \text{Part B } 
  \end{array} \right. \\
 &&(\rho_A^{i}, \phi_A^{i}[^{\circ}])=\left\{ \begin{array}{ll}
 & (2.85^{+2.18}_{-1.92}, -103^{+52}_{-63})\qquad   \text{Part ${\rm A}^{\prime}$ } \\
 & (6.54^{+1.81}_{-3.30}, -206^{+23}_{-24})\qquad   \text{Part ${\rm B}^{\prime}$ } 
  \end{array} \right. \\
 && (\rho_{H}, \phi_{H}[^{\circ}])=(3.09^{+1.64}_{-1.53}, -102^{+40}_{-31})\,.
 \label{soluSIII}
 \end{eqnarray}
 
 It can be easily seen from Fig.~\ref{SpecFig} that: (1) for factorizable annihilation 
 parameters $(\rho_A^{f}, \phi_A^{f})$, similar to scenarios I and II, there are two 
 allowed regions (labelled by part A and B);
 (2) for nonfactorizable annihilation parameters
 $(\rho_A^{i}, \phi_A^{i})$, besides the
 solution similar to scenarios I and II (labelled by part ${\rm A}^{\prime}$), 
 another solution (labelled by part ${\rm B}^{\prime}$) with a very large value 
 of $\rho_A^{i}$ is gotten. 
 (3) It is very intersting that the allowed space of $(\rho_{H}, \phi_{H})$ overlaps almost 
 entirely with the ``part ${\rm A}^{\prime}$'' allowed space of $(\rho_A^{i}, \phi_A^{i})$.
 Moreover, their best-fit points $(\rho_A^{i}, \phi_A^{i})$ $=$ $(2.85, -103^{\circ})$ of 
 ``part ${\rm A}^{\prime}$'' and $(\rho_{H}, \phi_{H})$ $=$ $(3.09, -102^{\circ})$ are
 very close to each other. It might imply that the assumption $X_A^{i}$ 
 $(\rho_A^{i}, \phi_A^{i})$ $=$ $X_H$ $(\rho_H, \phi_H)$ used in scenarios I and II 
 is a good simplification.
 
 With the best fit parameters in scenarios III, either the small value of $\rho_A^{i}$
 in ``part ${\rm A}^{\prime}$'' or the large value in ``part ${\rm B}^{\prime}$'',
 our evaluations on branching ratios, direct and mixing-induced $CP$ asymmetries
 for $B_{u,d}$ ${\to}$ $\pi K$, $K \bar{K}$, $\pi \pi$ decays are similar to those
 given in our scenarios I and II, so no longer listed here.
 For the two solutions ${\rm A}^{\prime}$ and  ${\rm B}^{\prime}$ 
 of $(\rho_A^{i}, \phi_A^{i})$, it is expected by QCDF approach \cite{Beneke2} 
 that the parameter $\rho_A^{i}$ should have a small value, which is also 
 favored by our scenarios I and II fit. In fact, such two solutions lead to the 
 same results of $A^{i}_{1,2}$, but the different ones of $A^{i}_{3}$, which 
 principally provides an opportunity to refute one of them. However, 
 because $A^{i}_{3}$ is numerically trivial due to 
 $(r_\chi^{M_1}-r_\chi^{M_2}) \sim 0$ for the light mesons, 
 such way is practically unfeasible for current accuracies of 
 theoretical calculation and experimentally measurement.
  
 \section{Conclusions}
 \label{sec04}
 The recent CDF and LHCb measurements of large branching
 ratios for pure annihilation $\bar{B}_s^0$
 $\to$ $\pi^+ \pi^-$ and $\bar{B}_d^0$ $\to$ $K^+ K^-$
 decays imply possible large annihilation contributions,
 which induce us to modify the traditional QCDF treatment
 for annihilation parameters.
 Following the suggestion of Ref.\cite{zhu2},
 two sets of annihilation parameters $X_A^i$ and $X_A^f$
 are used to parameterize the endpoint singularity
 in nonfactorizable and factorizable annihilation
 amplitudes, respectively.
 Besides annihilation effects, the resolution of
 so-called ${\pi}K$ and ${\pi}{\pi}$ puzzles also
 expect constructive contributions from
 spectator scattering topologies.
 With the approximation of $X_A^i$ $=$ $X_H$,
 we perform a global fit on both annihilation
 parameters ($\rho_{A}^{i,f}$, $\phi_{A}^{i,f}$)
 and $B$-meson wave function parameter $\lambda_B$
 based on available experimental data for $B$
 $\to$ $\pi \pi$, $\pi K$ and $K \bar{K}$
 decays. Our main conclusions and findings are
 summarized as:
 \begin{itemize}
 \item
 The $95\%$ C.L. allowed region of $(\rho_A^i,\phi_A^i)$
 is entirely different from that of $(\rho_A^f,\phi_A^f)$.
 This fact means that the traditional QCDF treatment
 $(\rho_A,\phi_A)$ as universal parameters for different
 annihilation topologies might be unapplicable to hadronic
 $B$ decays.
 \item
 The current experimental data on $B$ $\to$ $\pi \pi$,
 $\pi K$ and $K \bar{K}$ decays seems to favor a
 large value of $\rho_A^i$ $\sim$ 2.9, which corresponds
 to a sizable nonfactorizable annihilation contributions.
 But the range of $(\rho_A^i,\phi_A^i)$ is still very large,
 because the measurement precision of $CP$ asymmetries
 is low now.
 \item
 There are two possible choices for parameters $(\rho_A^f,\phi_A^f)$.
 One is $(\rho_A^f,\phi_A^f)$ $\sim$ $(1.1,-40^{\circ})$,
 the other is $(\rho_A^f,\phi_A^f)$ $\sim$ $(2.7,165^{\circ})$.
 These two choices correspond to similar factorizable annihilation
 contributions, although the QCDF approach tends to have
 a small value of $\rho_A^f$ \cite{Beneke2}.
 The space for $(\rho_A^f,\phi_A^f)$ is relatively
 tight due to the well measured branching ratios for
 $B$ $\to$ $\pi \pi$, $\pi K$ and $K \bar{K}$ decays.
 \item
 The spectator scattering corrections play an
 important role in resolving both $\pi K$ and $\pi \pi$
 puzzles. Within QCDF approach, the spectator
 scattering amplitudes depend on parameters
 $(\rho_H,\phi_H)$ and $B$-meson wave function
 parameter $\lambda_B$.
 In our analysis, the approximation $(\rho_H,\phi_H)$
 $=$ $(\rho_A^i,\phi_A^i)$ is assumed, 
 which is proven
 to be a good simplification by a global fit in scenario III. 
 A small value
 of $\lambda_B$ $\leq$ 350 MeV at the 95\% C.L. is
 obtained by the global fit on $B$ $\to$ $\pi \pi$,
 $\pi K$ and $K \bar{K}$ decays, which needs to
 be further tested by future improved measurement on
 $B$ $\to$ $\ell \nu_\ell \gamma$ decays.
 An enhanced color-suppressed tree coefficient
 $\alpha_2$, which is supported by both large value
 of $\rho_H$ $\sim$ 2.9 and small value of
 $\lambda_B$ $\sim$ 200 MeV, is helpful to
 reconcile discrepancies on $\Delta A$ and
 $R_{00}^{\pi \pi}$ between QCDF approach
 and experiments.
 \end{itemize}

 The spectator scattering and annihilation contributions
 can offer significant corrections to observables of
 hadronic $B$ decays, and deserve intensive research
 especially when we apply the QCDF approach to the
 penguin-dominated, color-suppressed tree, and pure
 annihilation nonleptonic $B$ decays.
 As suggested in Ref.\cite{zhu1,zhu2} and proofed by
 the pQCD approach \cite{pqcd}, different parameters
 corresponding to different topologies should be
 introduced to regulate the endpoint divergences in
 spectator scattering and annihilation amplitudes
 within QCDF approach,
 even parameters reflecting the flavor symmetry-breaking
 effects should be considered for $B_{u,d,s}$ decays
 \cite{zhu1,zhu2,Cheng1,Cheng2,Cheng3,Beneke2,chang1}.
 This treatment might provide possible solution to
 ``problematic'' discrepancies between QCDF results
 and available measurements.
 Of course, a fine-tuning of these parameters is
 required to be compatible with the experimental
 constraints.
 With the running LHCb and the upcoming SuperKEKB
 experiments, more refined measurements on $B$-meson
 decays can be obtained, which will provide more
 powerful grounds to test various approach and
 confirm or refute some theoretical hypotheses.

  \section*{Acknowledgments}
  \label{thanks}
  This work is supported by National Natural Science
  Foundation of China under Grant Nos. 11147008,
  11105043 and U1232101, 11475055. Q. Chang is also supported by
  Foundation for the Author of National Excellent Doctoral
  Dissertation of P. R. China under Grant No. 201317, 
  Research Fund for the Doctoral Program of Higher
  Education of China under Grant No. 20114104120002
  and Program for Science and Technology Innovation
  Talents in Universities of Henan Province (Grant No. 14HASTIT036).

 \begin{appendix}
 \section{Building blocks of annihilation and spectator scattering contributions}
 \label{app01}
 The annihilation amplitudes for two-body nonleptonic
 $B$ $\to$ $M_{1}M_{2}$ decays (here $M_{i}$ denotes the
 light pseudoscalar meson) can be expressed as the
 following building blocks \cite{Beneke2},
 \begin{eqnarray}
  A_1^i &=&
 \pi \alpha_s \int_0^1 dx dy
 \Big\{ \Phi_{M_2}^{a}(x) \Phi_{M_1}^{a}(y)
 \Big[ \frac{1}{y(1-x \bar{y})}
     + \frac{1}{\bar{x}^2 y} \Big]
     + r_\chi^{M_1} r_\chi^{M_2}
     \frac{2 \Phi_{M_2}^{p}(x) \Phi_{M_1}^{p}(y)}
          {\bar{x}y} \Big\}
 \label{ai1}, \\
 A_2^i &=&
 \pi \alpha_s \int_0^1  dx dy
 \Big\{ \Phi_{M_2}^{a}(x) \Phi_{M_1}^{a}(y)
 \Big[ \frac{1}{\bar x(1-x \bar{y})}
     + \frac{1}{\bar{x} y^2} \Big]
     + r_\chi^{M_1} r_\chi^{M_2}
     \frac{2 \Phi_{M_2}^{p}(x) \Phi_{M_1}^{p}(y)}
     {\bar{x} y} \Big\}
 \label{ai2}, \\
 A_3^i &=&
 \pi \alpha_s \int_0^1 dx dy
 \Big\{ r_\chi^{M_1}
    \frac{2 \bar{y}\ \Phi_{M_2}^{a}(x) \Phi_{M_1}^{p}(y)}
         {\bar{x}y(1-x\bar{y})}
      - r_\chi^{M_2}
    \frac{2x\ \Phi_{M_1}^{a}(y) \Phi_{M_2}^{p}(x)}
         {\bar{x}y(1-x\bar{y})} \Big\}
 \label{ai3}, \\
 A_1^f &=& A_2^f =0
 \label{af12}, \\
 A_3^f &=&
 \pi \alpha_s \int_0^1 dx dy
 \Big\{ r_\chi^{M_1}
    \frac{2(1+\bar x)\ \Phi_{M_2}^{a}(x) \Phi_{M_1}^{p}(y)}
         {\bar{x}^2 y}
      + r_\chi^{M_2}
    \frac{2(1+y)\  \Phi_{M_1}^{a}(y) \Phi_{M_2}^{p}(x)}
         {\bar{x} y^2} \Big\}
 \label{af3},
 \end{eqnarray}
 where the subscripts $k$ on $A^{i,f}_{k}$
 correspond to three possible Dirac current structures,
 namely, $k$ $=$ $1$, $2$, $3$ for $(V-A)\otimes(V-A)$,
 $(V-A)\otimes(V+A)$, $-2(S-P)\otimes(S+P)$,
 respectively.
 $r_\chi^{M}$ $=$ $2m_{M}^{2}/m_{b}(m_{1}+m_{2})$,
 where $m_{1,2}$ are the current quark mass of the
 pseudoscalar meson with mass $m_{M}$.
 $\Phi_{M}^{a}$ and $\Phi_{M}^{p}$ are the twist-2
 and twist-3 light-cone distribution amplitudes,
 respectively. Their asymptotic forms are
 $\Phi_{M}^{a}(x)$ $=$ $6x\bar{x}$ and
 $\Phi_{M}^{p}(x)$ $=$ $1$.

  The spectator scattering corrections are given by \cite{Beneke2}
  \begin{equation}
   H_i ( M_1 M_2) =
   \left\{ \begin{array}{l}
   \displaystyle
  +\frac{B_{M_1 M_2}}{A_{M_1 M_2}}
  {\int}_{0}^{1}d{\xi}
   \frac{ \Phi_{B}(\xi) }{ \xi }
   \int_0^1 dx dy \Big[
   \frac{ \Phi_{M_2}^{a}(x) \Phi_{M_1}^{a}(y) }
        { \bar{x} \bar{y} }
   + r_\chi^{M_1}
   \frac{ \Phi_{M_2}^{a}(x) \Phi_{M_1}^{p}(y) }
        { x \bar{y} } \Big],
   \\ \qquad \qquad \qquad
   \text{for }\, i=1,2,3,4,9,10
   \\
   \displaystyle
   -\frac{B_{M_1 M_2}}{A_{M_1 M_2}}
   {\int}_{0}^{1}d{\xi}
    \frac{ \Phi_{B}(\xi) }{ \xi }
    \int_0^1 dx dy \Big[
    \frac{ \Phi_{M_2}^{a}(x) \Phi_{M_1}^{a}(y) }
        { x \bar{y} }
   + r_\chi^{M_1}
   \frac{ \Phi_{M_2}^{a}(x) \Phi_{M_1}^{p}(y) }
        { \bar{x} \bar{y} } \Big],
   \\ \qquad \qquad \qquad
   \text{for }\, i=5,7
   \\
   0, \qquad \qquad \quad
   \text{for }\, i=6,8
  \end{array} \right.
  \label{hardblock}
  \end{equation}
 where the factorized matrix elements are
 parameterized as \cite{Beneke2}
 \begin{equation}
  A_{M_{1}M_{2}} =
  i\frac{G_{F}}{\sqrt{2}}
  m_{B}^{2}F_{0}^{B{\to}M_{1}}f_{M_{2}},
  \qquad \qquad
  B_{M_{1}M_{2}} =
  i\frac{G_{F}}{\sqrt{2}}
  f_{B}f_{M_{1}}f_{M_{2}}.
 \end{equation}

 \section{Theoretical input parameters}
 \label{app02}
 For the CKM matrix elements, we adopt the fitting
 results for the Wolfenstein parameters given by
 the CKMfitter group \cite{CKMfitter}
 \begin{eqnarray}
 \bar{\rho} = 0.140^{+0.027}_{-0.026}, \quad
 \bar{\eta} = 0.343^{+0.015}_{-0.014}, \quad
  A = 0.802^{+0.029}_{-0.011}, \quad
 \lambda = 0.22543^{+0.00059}_{-0.00094}.
 \end{eqnarray}

 The pole masses of quarks are \cite{PDG12}
 \begin{eqnarray}
 &&m_u=m_d=m_s=0, \quad
   m_c=1.67 \pm 0.07 \, {\rm GeV},
   \nonumber\\
 &&m_b=4.78 \pm 0.06 \, {\rm GeV}, \quad
    m_t=173.5 \pm 1.0\,{\rm GeV}
 \label{polemass}.
 \end{eqnarray}

 The running masses of quarks are \cite{PDG12}
 \begin{eqnarray}
 &&\frac{\bar{m}_s(\mu)}{\bar{m}_q(\mu)} = 27 \pm 1, \quad
   \bar{m}_{s}(2\,{\rm GeV}) = 95 \pm 5 \,{\rm MeV}, \quad
   \bar{m}_{c}(\bar{m}_{c}) = 1.275 \pm 0.025 \,{\rm GeV},
   \nonumber \\
 &&\bar{m}_{b}(\bar{m}_{b}) = 4.18 \pm 0.03 \,{\rm GeV}, \quad
   \bar{m}_{t}(\bar{m}_{t}) = 160.0^{+4.8}_{-4.3}\,{\rm GeV}
 \label{runningmass}.
 \end{eqnarray}

 The decay constants of $B$-meson and light mesons are \cite{PDG12}
 \begin{equation}
 f_{B} = (0.190 \pm 0.013)\,{\rm GeV}, \quad
 f_{\pi} = (130.4 \pm 0.2)\,{\rm MeV}, \quad
 f_{K} = (156.1 \pm 0.8)\,{\rm MeV}.
 \end{equation}

 We take the following heavy-to-light
 transition form factors \cite{BallZwicky}
 \begin{equation}
 F^{B \to \pi }_{0}(0) = 0.258 \pm 0.031, \qquad
 F^{B \to K }_{0}(0) = 0.331 \pm 0.041.
 \end{equation}

 Moreover, for the Gegenbauer coefficients,
 we take \cite{BallG}
 \begin{equation}
 a_{1}^{\pi}({\rm 2 GeV}) =0, \quad
 a_2^{\pi}({\rm 2 GeV}) =0.17, \quad
 a_{1}^{K}({\rm 2 GeV}) =0.05, \quad
 a_{2}^{K}({\rm 2 GeV}) =0.17.
 \end{equation}

 For the other inputs, such as the masses
 and lifetimes of mesons and so on,
 we take their central values
 given by PDG \cite{PDG12}.

 \section{Fitting Approach}
 \label{app03}
 Our fit is performed in a simple way, which is
 similar to the one adopted in Ref.\cite{Vernazza}
 based on the frequentist framework.
 Considering a set of $N$ observables $f_j$,
 the experimental measurements are assumed to
 be Gaussian distributed with the mean value
 $f_{j\,\rm exp}$ and error $\sigma_{j\,\rm exp}$.
 The theoretical prediction $f_{j\,\rm theo}$
 for each observable
 could be treated as a function of a set of
 ``unknown'' free parameters
 $\{y_i\}$ (here $y_i$ $=$ $\rho_{A}^{i,f}$,
 $\phi_{A}^{i,f}$ and $\lambda_B$ in this paper).
 To estimate the values of ``unknown'' parameters
 $\{y_i\}$ and compare the theoretical results
 $f_{j\,\rm theo}$ with the experimental data
 $f_{j\,\rm exp}$, typically, it is need to
 construct a $\chi^2$ function as
 \begin{equation}
 \chi^2(\{y_i\}) =
 \sum\limits_{j=1}^N
 \frac{(f_{j\,\rm theo}(\{y_i\})-f_{j\,\rm exp})^2}
      {\sigma_{j\,\rm exp}^2}.
 \label{chi2I}
 \end{equation}

 In the evaluation of $f_{j\,\rm theo}$ for
 hadronic B decays, ones always encounter
 theoretical uncertainties induced by input
 parameters, like form factor and decay
 constant, whose probability distribution is
 unknown.
 Following the treatment of Rfit scheme
 \cite{CKMfitter, Rfit} that input values are
 treated on an equal footing, irrespective of
 how close they are from the edge of the
 allowed range, the $\chi^2$ function is
 modified as \cite{Vernazza}
 \begin{equation}
 \chi^2 = \sum_{j=1}^N
 \left\{ \begin{array}{cl}
 \displaystyle
 \frac{([ f_{j\,\rm theo}-\delta_{j\,\rm theo,\,sub} ]
       -f_{j\,\rm exp})^2}{\sigma_{j\,\rm exp}^2}
  & \quad \text{if } f_{j\,\rm exp} <
   [f_{j\,\rm theo}-\delta_{j\,\rm theo,\,sub}],
 \\
 \displaystyle
 \frac{(f_{j\,\rm exp}-
   [f_{j\,\rm theo}+\delta_{j\,\rm theo,\,sup}])^2}
     {\sigma_{j\,\rm exp}^2}
 & \quad \text{if  } f_{j\,\rm exp} >
  [f_{j\,\rm theo}+\delta_{j\,\rm theo,\,sup}],
 \\
 0 & \quad \text{otherwise}
 \end{array} \right.
 \label{chi2II}
 \end{equation}
 where $\delta_{j\,\rm theo,\,sup}$ and
 $\delta_{j\,\rm theo,\,sub}$ denote asymmetric
 theoretical uncertainties, and are defined as
 $(f_{j\,\rm theo})^{+\delta_{j\,\rm theo,\,sup}}_{-\delta_{j\,\rm theo,\,sub}}$.
 As to the asymmetric experimental errors,
 we choose the larger one as weighting factor.
 Correspondingly, the confidence levels are
 defined by the function
 \begin{equation}
 {\rm CL}(\{y_i\}) = \frac{1}{\sqrt{2^{N_{\rm dof}}} \Gamma(N_{\rm dof}/2)}
  \int_{\Delta \chi^2(\{y_i\})}^{\infty} e^{-t/2}t^{N_{\rm dof}/2 -1}dt
  \label{CLfun},
 \end{equation}
 with $\Delta \chi^2$ $=$ $\chi^2$ $-$ $\chi^2_{\rm min}$
 and $N_{\rm dof}$ the number of degrees of freedom
 of free parameters.

 With the input parameters summarized in
 Appendix \ref{app02}, we scan the space of the
 parameters $y_i$ and calculate the theoretical
 results $f_{j\,\rm theo}$.
 The $\chi^2 $ could be obtained with Eq.(\ref{chi2II}).
 The numerical results at $1 \sigma$ and $2 \sigma$
 confidence levels are gotten from Eq.(\ref{CLfun})
 by taking ${\rm CL}$ $=$ $1 - 68.27\%$ and
 ${\rm CL}$ $=$ $1 - 95.45\%$, respectively.

\end{appendix}

 \end{document}